\documentclass[
reprint,
superscriptaddress,
showkeys,
amsmath,
amssymb,
aps,
floatfix,
]{revtex4-2}

\usepackage{graphicx}
\usepackage{dcolumn}
\usepackage{bm}
\usepackage{float}
\usepackage[english]{babel} 
\usepackage{tabularray}
\usepackage{array}
\usepackage{natbib}[sort&compress]
\usepackage{xcolor}
\usepackage{color}
\usepackage{soul}
\usepackage{amssymb, nicefrac}
\usepackage[most]{tcolorbox}
\usepackage{hyperref}
\hypersetup{
	colorlinks=true,
	linkcolor=blue,
	filecolor=magenta,      
	urlcolor=blue,
	citecolor=blue,
	pdfpagemode=FullScreen,
}
\usepackage{multirow}
\usepackage{cleveref}
\crefname{figure}{Fig.}{Figs.}
\crefname{equation}{Eq.}{Eqs.}

\begin{document}
	
	
	\title{ESTIMATION OF CORRELATION COEFFICIENTS AND SPIN ANGULAR DISTRIBUTIONS OF FISSION FRAGMENTS}
	
	\author{D.E. Lyubashevsky}
	\email[]{lyubashevskiy@phys.vsu.ru} 
	\affiliation{Voronezh State University, Voronezh, Russia}
	\affiliation{International Institute of Computer Technologies, Voronezh, Russia}
	\author{A.A. Pisklyukov}
	\author{S.V. Klyuchnikov}
	\affiliation{Voronezh State University, Voronezh, Russia}
	\author{P.V. Kostryukov}
	\affiliation{Voronezh State University of Forestry and Technologies, Voronezh, Russia}
	\affiliation{Voronezh State University, Voronezh, Russia}
	
	\date{\today}
	
	\begin{abstract}
		This study proposes a theoretical model for studying the spin characteristics and angular correlations of fission fragments of heavy nuclei. The mechanisms of spin formation, including the influence of transverse vibrations, are considered and the relationship between the anisotropy of the angular distribution and the correlation coefficient is revealed. The theoretical predictions are compared with experimental data and various models developed by other research groups. 
	\end{abstract}
	
	\maketitle
	
	\section{Introduction}
	
	The study of the properties of nuclear fission products, particularly their spin characteristics, is a key area in nuclear physics. Understanding these characteristics provides deeper insights into the fundamental processes occurring during the fission of heavy nuclei. One of the central questions is the mechanism of spin formation in fission fragments (FF) and the potential correlations between them. According to quantum fission theory, the spin distributions (SD) of primary FFs determine the dynamical properties of the fission process, including the multiplicities, energy, and angular distributions of evaporated and delayed neutrons, as well as gamma rays. Despite many years of research, there is still no unified theory capable of explaining all the experimentally observed data.
	
	There has been particular interest in studying the SDs of fission products from actinide nuclei, such as Th, U, and Cf. A significant contribution in this area was made by the collaboration in~\cite{wilson2021}, which provided experimental values for the spin characteristics for low-energy induced fission of $^{232}$Th and $^{238}$U, as well as spontaneous fission of $^{252}$Cf. Based on these results, the researchers reached the unexpected conclusion that the spins of light and heavy FFs are uncorrelated, contradicting previous assumptions that the fragment spins should be linked at the moment of fission. They proposed a model in which spins form at a late stage of the fission process: after the nucleus has split into fragments but before the emission of neutrons and gamma rays. This finding represents a significant advance in understanding the fission mechanism but requires further theoretical and experimental validation.
	
	However, the theoretical description of the formation of FF spins remains challenging. Most modern models~\cite{randrup2021,bulgac2021,rasmussen1969} describing SDs are based on transverse vibrational modes of the fissile nucleus that occur before it reaches the scission point. These modes, first considered in~\cite{nix1965}, include bending, wriggling, twisting, and tilting, each corresponding to different types of collective motion of the pre-fragments. Among these, the bending and wriggling modes are particularly important for understanding spin generation. In the bending mode, one pre-fragment rotates clockwise or counterclockwise around an axis perpendicular to the symmetry axis of the fissioning nucleus, while the other pre-fragment rotates in the opposite direction. This motion results in a bending deformation at the neck, and the sum of the spins generated by these vibrations must be zero due to the total spin conservation law. In the wriggling mode, however, both pre-fragments rotate in the same direction around the same axis, resulting in large, aligned spins. To conserve the total spin of the fissioning system, the nucleus must rotate in the opposite direction, generating significant orbital angular momentum of the fragments. These vibrational modes are critical for understanding the dynamics of spin generation and the angular momentum distribution in fission fragments.
	
	Despite these models' success in explaining the generation of large FF spins $J_1$ and $J_2$ ($\approx 5-7\hbar$) relative to the spin of the compound nucleus $J$ ($\approx 1\hbar$), they cannot fully account for the lack of spin correlation. For instance, the Randrup model~\cite{randrup2021} predicts no correlations between the spins of FFs. However, the use of transverse bending and wriggling vibrations in this model implies that the FF spins are formed before the rupture of the fissile system, which contradicts the spin formation mechanism proposed in~\cite{wilson2021} and the stage of fission at which these spins are formed. This discrepancy may arise from an inaccurate description of the fission dynamics, such as the contributions of the aforementioned vibrations or fragment deformations.
	
	Therefore, the present work aims to develop a theoretical model within the framework of quantum fission theory to explain the experimentally observed SDs of fragments~\cite{wilson2021}. The resulting distributions of this model will be compared with experimental data to search for correlations in the SDs. The results will also be compared with theoretical data obtained by groups using different approaches. Special attention will be given to analyzing the angular distributions (ADs) of the FF spins, which will be compared using two different approaches.
	
	The first approach employs a temperature-based model that takes into account bending and wriggling modes within a two-dimensional spin model~\cite{randrup2021,vogt2021}. The second approach proposes a three-dimensional model by adding longitudinal modes, such as twisting and tilting, and calculates the spin characteristics within the framework of time-dependent density functional theory (TDDFT)~\cite{stetcu2021}. This work also aims to investigate whether the experimentally determined~\cite{wilson2021} absence of FF spin correlations can shed light on the stage at which FF spins form.
	
	These studies will not only help to verify existing models but also suggest new approaches to describe the fission processes of heavy nuclei. This may lead to a better understanding of the fission mechanism and spin formation, thereby advancing the theory of nuclear fission.

	\section{Spin distributions modelling}
	
	In the framework of the developed approach~\cite{kadmensky2002, kadmensky2003, kadmensky2005, kadmensky2008, kadmensky2009}, the spin formation process is determined by the behavior of the compound nucleus from the outer saddle point to the scission point~\cite{kadmensky_yad2024}. This process is described using the harmonic oscillator approximation, where the spins of the pre-fragments are related solely to their momentum distribution, as proposed in~\cite{moretto1980}. As a result, the SDs of the fragments are constructed~\cite{nix1965, kadmensky2017, kadmensky_yad2024} using zero oscillatory wave functions in the momentum representation along the $X$ and $Y$ axes.
	
	The theoretical foundation of this approach is rooted in the concept of transitional fission states, first introduced in~\cite{bohr1939}. In this model, the fission process is described as a transition through a series of quasi-stationary states near the saddle point of the nuclear potential. The mentioned works~\cite{kadmensky2002, kadmensky2003, kadmensky2005, kadmensky2008, kadmensky2009} extended this idea by showing that out of the total number of quasiparticle states on the order of $10^6$, only about $10^2$ correspond to configurations that lead to fission. These states are characterized by the nucleus being located beyond the outer saddle point, where the majority of the excitation energy is converted into the energy of non-equilibrium deformation of the FFs, rather than thermal excitation. This phenomenon is referred to as the ``coldness'' of the fissioning nucleus.
	
	The ``cold'' nature of the fissioning system ensures that it retains axial symmetry and preserves the projection $K$ of the spin $J$ on the symmetry axis, as discussed in~\cite{bohr1998}. Transitional states formed at the saddle points of the deformation potential act as filters, selecting the most probable values of $K$, which determine the characteristics of partial fission widths and ADs of the FFs. The preservation of $K$ is crucial for explaining the observed anisotropies in the ADs of fission products, as it prevents the mixing of spin projections that would occur in a heated system.
	
	Experimental data on low-energy fission, including reactions with polarized neutrons and gamma quanta~\cite{danilyan2019, gagarski2016}, show anisotropies in the ADs of fission products. These anisotropies, characterized by the coefficient $A \approx 0.1$ for actinide nuclei, confirm that the fissioning system remains ``cold'' at all stages of fission. The preservation of the $K$ projection is consistent with our model, as heating the nucleus would lead to uniform mixing of the $K$ projections and the disappearance of anisotropies.
	
	The anisotropy of the ADs, observed in experiments on neutron~\cite{danilyan2009, danilyan2010, danilyan2019} and gamma-quanta~\cite{valsky2010, gagarskii2011} induced fission, can be described by the following form:
	\begin{equation}
		P(\varphi) \sim 1 + A \cos^2{\varphi},
	\end{equation}
	where $\varphi$ is the angle between the evaporative neutron (or gamma quantum) flight direction and the light FF flight direction.
	
	The description of this anisotropy is based on the assumption that the FF spins are oriented perpendicular to the direction of the light fragment's motion, as proposed in~\cite{ericson1958}. This assumption is supported by experimental data from~\cite{wilhelmy1972, wolf1976}, where the ADs of fission products were found to be consistent with the spins being perpendicular to the fission direction. If the spin had a non-zero projection along the $Z$ axis, this would reduce the perpendicular components, leading to a decrease in the observed asymmetry coefficients. However, since no such reduction is observed experimentally, it is concluded that the spin projection on the $Z$ axis is zero. This allows us to use a two-dimensional model of spins, consistent with experimental data~\cite{wilson2021}.
	
	The probability distribution of the spins of two independent vibrations, i.e., the mentioned transverse wriggling and bending ones, can be represented as the product of the distributions~\cite{kadmensky2017, kadmensky_yad2024}:
	\begin{equation}\label{eq:1}
		P \! \left( \! J_{k_x}, \! J_{k_y} \! \right) \! \equiv \! P \! \left( \! J_{k_x} \right) \! P \! \left( \! J_{k_y} \right) \! = \! \frac{1}{ \pi C_k} \! \exp \! \left[ \!  - \frac{J_{k_x}^2 \! + \! J_{k_y}^2}{C_k} \right] \! ,
	\end{equation}
	where the index $k={\rm w},b$ corresponds to the type of oscillation, the value $C_k$ is related to the moments of inertia $I_k$ and oscillation frequencies $\omega_k$ by the relation $C_k = I_k \hbar \omega_k$. The frequencies are determined by the classical formulas:
	$$
	\omega_k=\sqrt{\frac{K_k}{I_k}},
	$$
	where $K_k$ is the stiffness coefficient~\cite{nix1965} and $I_k$ is the moment of inertia of the FF for the present type of vibration. Following the same ideas of the mentioned paper, we assume that $\hbar \omega_{\rm w} =2.5$ MeV and $\hbar \omega_b=0.9$ MeV, respectively, while admitting that the variation of the value depends weakly on the nucleon composition of the nucleus.
	
	The moment of inertia of the wriggling vibrations is related to the FF moments of inertia $I_1$ and $I_2$ by the representations of the works~\cite{randrup2021,randrup_dossing2022,vogt2021} as follows:
	\begin{equation}\label{eq:2}
		I_{\rm w} =\frac{(I_1+I_2)I_0}{I},
	\end{equation}
	where $I = I_0 + I_1 + I_2$ is the total moment of inertia, and $I_0=\frac{M_1 M_2}{M_1 + M_2} {\left( R_1 + R_2 + d \right)}^2$ is the moment of inertia of the nucleus. The radii $R_i$ are determined by the corresponding deformation parameters $\beta_i$ by the relation:
	$$
	R_i=r_0 A^{1/3} \left[ 1-\beta_i^2/4\pi + \beta_i \sqrt{5/4\pi}\, \right],
	$$
	with the parameter $d\approx 4$ fm. The moment of inertia of the bending vibrations $I_b$ is determined~\cite{shneidman2002} by the relation:
	\begin{equation}\label{eq:3}
		I_b = I_1 + {\left(\frac{R_1}{R_2}\right)}^2 I_2.
	\end{equation}
	Note that, in such studies, the moments of inertia of the fragments are usually calculated within the framework of the rigid body model, i.e., $I_{1,2}\equiv I_{i_{rigid}}=\frac{M_i}{5}\sum{R_i^2}$. However, our method uses the approach of the Migdal superfluid nucleus model~\cite{migdal1960}, in which the moments of inertia of the FF differ significantly from their rigid body counterparts. Typically, the moments of inertia of strongly deformed FFs range from $0.4I_{rigid}$ to $0.7I_{rigid}$, and in the region of near-magic nuclei, the range is from $0.2I_{rigid}$ to $0.3I_{rigid}$. For details on this matter, the authors refer to the following paper~\cite{lyubashevsky2025}, where a more detailed study has been performed. Using the estimation of the FF moments of inertia from~\cite{migdal1960} and the formulas~\cref{eq:2,eq:3}, one can obtain the indicated values for the wriggling and bending vibrations, which are given in~\Cref{tab:t1} for the cases of induced fission of nuclei $\rm ^{232}Th$, $\rm ^{238}U$, and spontaneous fission of $\rm ^{252}Cf$.
	
	Analyzing the data presented in~\Cref{tab:t1}, it can be seen that the moments of inertia of FF differ significantly depending on the type of oscillation. For example, for $\rm Xe$ and $\rm Ba$ nuclei, the values of the moments of inertia $I_b$ are higher than for wriggling vibrations, and for some, such as $\rm Nd$, the opposite is true—high values of $I_{\rm w}$, especially noticeable in the fission reaction $\rm ^{252}Cf(sf)$. This indicates that different modes of oscillation have different effects on the dynamics of the formation of FF spins, which emphasizes the complexity of the processes occurring in nuclear fission. The necessity to take into account different vibration mechanisms when analyzing the spin characteristics of FFs is in excellent agreement with the theoretical ideas of the author's group~\cite{kadmensky2017,kadmensky_yad2024}.
	
	\renewcommand{\arraystretch}{1.25}
	\begin{table}
		\caption{\label{tab:t1}FF's moments of inertia for wriggling and bending vibrations for the studied reactions, with dimensionality $\rm \hbar \cdot MeV^{-1}$}
		\begin{ruledtabular}
			\begin{tabular}{c|cc|cc|cc}
				\multirow{2}{*}{Fragment} &
				\multicolumn{2}{c|}{$\rm ^{232}Th \left( n, f \right)$} &
				\multicolumn{2}{c|}{$\rm ^{238}U \left( n, f \right)$}  &
				\multicolumn{2}{c}{$\rm ^{252}Cf \left( sf \right)$}  \\
				&
				$I_{\rm w}$ & $I_b$ & $I_{\rm w}$ & $I_b$ & $I_{\rm w}$ & $I_b$ \\
				\hline
				$\rm ^{82}Ge$  & 35.86 & 27.28 & 27.59 & 37.96 &       &       \\
				$\rm ^{84}Ge$  & 36.08 & 30.66 &       &       &       &       \\
				$\rm ^{84}Se$  & 35.92 & 29.19 & 29.55 & 37.99 &       &       \\
				$\rm ^{86}Se$  & 35.49 & 31.76 & 31.59 & 38.05 &       &       \\
				$\rm ^{88}Se$  & 35.56 & 33.60 & 33.42 & 38.08 &       &       \\
				$\rm ^{88}Kr$  & 33.97 & 26.45 & 26.31 & 36.47 &       &       \\
				$\rm ^{90}Kr$  & 35.23 & 34.46 & 34.07 & 37.86 &       &       \\
				$\rm ^{92}Kr$  & 35.37 & 37.32 & 37.34 & 37.58 &       &       \\
				$\rm ^{94}Kr$  & 35.34 & 40.67 & 40.01 & 37.86 &       &       \\
				$\rm ^{92}Sr$  & 33.01 & 27.26 &       &       &       &       \\
				$\rm ^{94}Sr$  & 22.83 & 24.53 & 38.71 & 37.54 & 38.94 & 42.76 \\
				$\rm ^{96}Sr$  & 23.26 & 27.07 & 42.08 & 37.81 & 42.12 & 43.07 \\
				$\rm ^{98}Sr$  & 23.52 & 28.05 & 43.44 & 37.63 & 43.22 & 42.93 \\
				$\rm ^{98}Zr$  & 23.49 & 27.75 & 43.34 & 37.44 & 42.90 & 42.86 \\
				$\rm ^{100}Zr$ & 23.88 & 28.81 & 45.30 & 37.28 & 44.37 & 42.82 \\
				$\rm ^{102}Zr$ &       &       & 29.87 & 25.01 & 45.69 & 42.61 \\
				$\rm ^{104}Zr$ &       &       & 30.73 & 25.23 & 46.94 & 42.18 \\
				$\rm ^{102}Mo$ &       &       & 29.11 & 24.62 & 44.15 & 42.20 \\
				$\rm ^{104}Mo$ &       &       & 30.58 & 25.26 & 46.49 & 42.44 \\
				$\rm ^{106}Mo$ &       &       &       &       & 47.29 & 41.58 \\
				$\rm ^{108}Mo$ &       &       &       &       & 47.33 & 41.25 \\
				$\rm ^{108}Ru$ &       &       &       &       & 45.67 & 40.83 \\
				$\rm ^{110}Ru$ &       &       &       &       & 30.23 & 36.68 \\
				$\rm ^{112}Ru$ &       &       &       &       & 30.92 & 36.06 \\
				$\rm ^{112}Pb$ &       &       &       &       & 22.53 & 24.32 \\
				$\rm ^{114}Pb$ &       &       &       &       & 23.39 & 24.03 \\
				$\rm ^{116}Pb$ &       &       &       &       & 33.10 & 34.97 \\
				$\rm ^{130}Sn$ & 24.49 & 23.32 & 24.49 & 25.71 & 33.47 & 25.81 \\
				$\rm ^{132}Sn$ & 24.15 & 23.21 & 24.28 & 25.44 & 28.38 & 27.20 \\
				$\rm ^{134}Sn$ &       &       & 24.39 & 25.15 &       &       \\
				$\rm ^{132}Te$ & 24.07 & 23.29 & 24.17 & 25.55 &       &       \\
				$\rm ^{134}Te$ & 23.70 & 23.48 & 24.21 & 25.33 & 37.12 & 24.61 \\
				$\rm ^{136}Te$ & 23.42 & 23.58 & 24.24 & 25.11 & 38.36 & 24.24 \\
				$\rm ^{138}Te$ & 22.87 & 23.99 & 24.35 & 24.82 &       &       \\
				$\rm ^{138}Xe$ & 23.00 & 23.84 & 24.43 & 24.74 & 38.38 & 24.03 \\
				$\rm ^{140}Xe$ & 35.03 & 42.82 & 42.81 & 37.29 & 58.31 & 36.06 \\
				$\rm ^{142}Xe$ & 35.29 & 44.91 & 44.53 & 37.76 & 60.26 & 36.68 \\
				$\rm ^{142}Ba$ & 35.27 & 44.43 & 44.31 & 37.56 & 46.59 & 40.88 \\
				$\rm ^{144}Ba$ & 35.53 & 46.12 & 45.85 & 37.86 & 48.14 & 41.25 \\
				$\rm ^{146}Ba$ & 35.46 & 47.45 & 47.52 & 37.43 & 48.77 & 41.58 \\
				$\rm ^{148}Ba$ &       &       & 50.11 & 37.89 &       &       \\
				$\rm ^{148}Ce$ & 36.03 & 48.59 & 49.23 & 37.73 & 50.05 & 42.18 \\
				$\rm ^{150}Ce$ & 36.27 & 50.64 & 51.04 & 38.05 & 51.65 & 42.61 \\
				$\rm ^{152}Nd$ &       &       &       &       & 53.08 & 42.74 \\
				$\rm ^{154}Nd$ &       &       &       &       & 53.92 & 42.93 \\
			\end{tabular}
		\end{ruledtabular}
	\end{table}
	
	Now let us write out the relation between the spins of the wriggling and bending vibrations $J_k$ and the spins of the fragments $J_i$, taking into account the moments of inertia of the FF~\cite{shneidman2002, randrup2021, randrup_dossing2022, vogt2021} for the projections on the $X$ and $Y$ axes perpendicular to the symmetry axis of the fissile nucleus $Z$, which are of the form:
	\begin{equation}\label{eq:4}
		J_{i_p}=\frac{I_i}{I_1 + I_2} J_{\text{w}_p} + (-1)^{i+1} J_{b_p},
	\end{equation}
	where $p = x, y$. 
	Using this formula, let us express the spin projections of the wriggling vibrations by the spin projections of the fragments:
	\begin{equation}\label{eq:5}
		J_{{\rm w}_p} = J_{1_p}+J_{2_p}.
	\end{equation}
	Similarly, the projection of spin for bending vibrations, represented by $J_{b_p}$, is defined in terms of the following relations:
	\begin{equation}\label{eq:6}
		J_{b_{p}} = J_{1_p} - \frac{I_1}{I_1 + I_2} J_{\text{w}_p} = \frac{I_2  J_{1_p} - I_1  J_{2_p}}{I_1 + I_2}.
	\end{equation}
	As demonstrated in~\cite{kadmensky2017,kadmensky_yad2024}, the application of the formula~\cref{eq:1} and the transition from the distribution of the spins of the vibrations to the analogous distribution of the spins of the fragments, according to~\cref{eq:4,eq:5,eq:6}, with consideration of the transition Jacobian, yields the following result:
	\begin{widetext}
		\begin{equation}\label{eq:7}
			\begin{aligned}
				P \! & \left(J_{1_x}\!,J_{2_x}\!, J_{1_y}\!, J_{2_y} \right) \! \equiv \! P(J_{\text{w}_x}) P(J_{\text{w}_y}) P(J_{b_x}) P(J_{b_y})  \left| \! \frac{\partial P \left(J_{{\rm w}_x}, J_{b_x}, J_{{\rm w}_y}, J_{b_y} \right)}{\partial P \left(J_{1_x}, J_{2_x}, J_{1_y}, J_{2_y} \right) } \! \right| \! = \!
				\frac{1}{\pi^2 C_b C_{\rm w}} \exp \! \left[\! - \! \left( \frac{J_{{\rm w}_x}^2 + J_{{\rm w}_y}^2}{C_{\rm w}} \! + \! \frac{J_{b_x}^2 + J_{b_y}^2}{C_b} \right) \right] \! \times \\
				& \! \times \left| \! \frac{\partial P \left(J_{{\rm w}_x}, J_{b_x}, J_{{\rm w}_y}, J_{b_y} \right)}{\partial P \left(J_{1_x}, J_{2_x}, J_{1_y}, J_{2_y} \right) } \! \right| \! = \!
				\frac{1}{\pi^2 C_{\rm w} C_b} \exp \! \left[-\frac{(J_{1_x} \! + \! J_{2_x})^2 \! + \! (J_{1_y} \! + \! J_{2_y})^2}{C_{\rm w}} - \frac{(I_2 J_{1_x} \! +\! I_1 J_{2_x})^2 \! +\! (I_2 J_{1_y} \! + \! I_1 J_{2_y})^2}{C_b (I_1 \! + \! I_2)^2}\right]\!.
			\end{aligned}
		\end{equation}
	\end{widetext}
	After transitioning to the spherical coordinate system, the azimuthal angles, represented by the symbols $\varphi_1$ and $\varphi_2$, are replaced by the angles $\varphi'= (\varphi_1 + \varphi_2)/2$ and $\varphi = \varphi_1 - \varphi_2$. Subsequently, integration is performed over the angle $\varphi'$ within the interval $0 \le {\varphi}' \le 2\pi$, yielding the dependence of SD on the angle $\varphi$ within the interval $0 \le \varphi \le \pi$ between the FF spins. This dependence can be expressed in a simple analytical form:
	\begin{equation}\label{eq:8}
		\begin{aligned}
			P(J_1, J_2, \varphi) \! = \! \frac{2 J_1 J_2}{\pi C_{\rm w} C_b} \exp \! \left[-J_1^2(\alpha I_2^2  + \beta) \right. \\ \left. -  J_2^2 (\alpha I_{1}^{2} + \beta) + 2J_1 J_2 \cos \! \varphi (\alpha I_1 I_2 - \beta) \right],
		\end{aligned}
	\end{equation}
	where $\alpha = C_b^{-1}(I_1 + I_2)^{-2}$ and $\beta = C_{\rm w}^{-1}$.
	
	By determining the type of SD, it is possible to ascertain the mean values of spins and standard deviations, as well as a variety of other significant quantities. As previously stated in the Introduction, researchers of the paper~\cite{wilson2021} were able to experimentally establish the correlation coefficient between the spins. The obtained result, as defined by~\eqref{eq:7}, allows us to calculate the expected theoretical value of this quantity and compare it with the observed experimental result. Furthermore, it can be compared with estimates from other theoretical approaches, e.g.,~\cite{randrup2021}, which will be discussed in the following section of this paper.
	
	\section{Determination of spin correlation coefficients of fragments}
	
	In the analysis of systems comprising two or more random variables, the correlation moment and correlation coefficient offer convenient statistical quantities. The correlation moment, denoted $\mu_{J_1 J_2}$, is the mathematical expectation of the product of the deviations of the random variables of spins $J_1$ and $J_2$ as
	\begin{equation}\label{eq:9}
		\begin{aligned}
			\mu_{J_1 J_2} = \int\limits_0^\infty \int\limits_0^\infty & \left(J_1 - \langle J_1 \rangle \right) \left( J_2 - \langle J_2 \rangle \right) \times \\
			& \times P \left( J_1, J_2 \right) dJ_1 dJ_2,
		\end{aligned}
	\end{equation}
	where $P\left( J_1,J_2 \right)$ is the probability density distribution function of the spin probability densities of the two FFs, independent from the angle $\varphi$.
	
	To obtain this, one must integrate~\cref{eq:9} over the given angle $\varphi$, which can be done explicitly by using the relation:
	\begin{equation}\label{eq:10}
		\int\limits_0^\pi e^{x \cos \! \varphi}  d\varphi = \pi \tilde{j}_0 (t)
		= \pi j(it) = \pi \sum\limits_{k=0}^{\infty} \frac{x^{2k}}{\left(2^k k!\right)^2},
	\end{equation}
	in which $j(it)$ denotes the zero-order Bessel function of the first kind, and $\tilde{j}_0(t)$ is the modified zero-order Bessel function of the first kind.
	
	Following this operation, the probability distribution $P(J_1, J_2)$ is as follows:
	\begin{equation}\label{eq:11}
		\begin{aligned}
			P(J_1,J_2) = \frac{2 J_1 J_2}{C_{\rm w} C_b} \sum\limits_{n=0}^{\infty} {\left[ \frac{J_1 J_2}{n!}(\alpha I_1 I_2 - \beta)^n \right]^2} \times \\ 
			\times \left[ -J_1^2(\alpha I_2^2 + \beta) - J_2^2(\alpha I_1^2 + \beta)\right].
		\end{aligned}
	\end{equation}
	The correlation coefficient $c_{J_1 J_2} \left(A_1,A_2 \right)$ of the spins of fragments $J_1$ and $J_2$ represents the ratio of the correlation moment to the product of the standard deviations $\sigma_{J_1}$ and $\sigma_{J_2}$. These quantities are defined as:
	\begin{equation}\label{eq:12}
		c_{J_1 J_2} \left(A_1,A_2 \right) = \frac{\mu_{J_1 J_2}}{\sigma_{J_1} \sigma_{J_2}},
	\end{equation}
	where $\sigma_{J_i}=\sqrt{\langle J_i^2\rangle - \langle J_i\rangle^2}$. 
	
	The coefficient $c_{J_1 J_2} \left(A_1,A_2 \right)$ for each pair of mass numbers $A_1$ and $A_2$ is calculated individually, and the generalized coefficient $\tilde{c}_{J_1 J_2}$ is of the form:
	\begin{equation}\label{eq:13}
		\tilde{c}_{J_1 J_2} = \frac{\sum c_{J_1J_2} \left(A_1,A_2 \right) Y \left(A_1,A_2\right)}{\sum Y \left(A_1,A_2 \right)},
	\end{equation}
	where $Y\left(A_1,A_2 \right)$ is the yield for a given pair of fragments having mass numbers $A_1$ and $A_2$.
	
	Correlation coefficients calculated by~\cref{eq:12,eq:13} for both fragments and generalized form for the investigated reactions $\rm ^{232}Th(n,f)$, $\rm ^{238}U(n,f)$, and $\rm ^{252}Cf(sf)$ are presented in \Cref{tab:t2}. It should be noted that the values of $\tilde{c}_{J_1 J_2}$ show a complete absence of correlations between the FF spins, i.e., it can be stated that the proposed theoretical model within the framework of the ``cold'' nucleus approach is in good agreement with the experimentally obtained data~\cite{wilson2021}.
	
	\begin{table}
		\caption{\label{tab:t2}Coefficients $c_{J_1 J_2}$ and $\tilde{c}_{J_1J_2}$, and yields $Y \! (A_f)$ for the studied reactions}
		\begin{ruledtabular}
			\begin{tabular}{c|cc|cc|cc}
				\multirow{2}{*}{Fragment} &
				\multicolumn{2}{c|}{$\rm ^{232}Th \left( n, f \right)$} &
				\multicolumn{2}{c|}{$\rm ^{238}U \left( n, f \right)$}  &
				\multicolumn{2}{c}{$\rm ^{252}Cf \left( sf \right)$}  \\
				&
				$c_{J_1J_2}$ & $Y(A_f)$ & $c_{J_1J_2}$ & $Y(A_f)$ & $c_{J_1J_2}$ & $Y(A_f)$ \\
				\hline
				$\rm ^{82}Ge$  & 0.203 & 0.64 & 0.207 & 0.12 &       &      \\
				$\rm ^{84}Ge$  & 0.196 & 0.32 &       &      &       &      \\
				$\rm ^{84}Se$  & 0.198 & 1.09 & 0.202 & 0.17 &       &      \\
				$\rm ^{86}Se$  & 0.085 & 4.68 & 0.121 & 0.84 &       &      \\
				$\rm ^{88}Se$  & 0.042 & 2.21 & 0.064 & 0.54 &       &      \\
				$\rm ^{88}Kr$  & 0.114 & 0.85 & 0.133 & 0.37 &       &      \\
				$\rm ^{90}Kr$  & 0.027 & 5.34 & 0.017 & 1.85 &       &      \\
				$\rm ^{92}Kr$  & 0.012 & 3.92 & 0.005 & 2.50 &       &      \\
				$\rm ^{94}Kr$  & 0.001 & 0.60 &       &      &       &      \\
				$\rm ^{92}Sr$  & 0.062 & 0.20 & 0.006 & 0.73 &       &      \\
				$\rm ^{94}Sr$  & 0.007 & 2.04 & 0.006 & 1.51 & 0.055 & 0.55 \\
				$\rm ^{96}Sr$  & 0.020 & 3.54 & 0.002 & 4.13 & 0.018 & 0.89 \\
				$\rm ^{98}Sr$  & 0.019 & 1.32 & 0     & 2.27 & 0.015 & 0.37 \\
				$\rm ^{98}Zr$  & 0.020 & 0.37 & 0     & 0.49 & 0.014 & 0.59 \\
				$\rm ^{100}Zr$ & 0.047 & 0.88 & 0.016 & 3.30 & 0.001 & 2.06 \\
				$\rm ^{102}Zr$ &       &      & 0.033 & 4.09 & 0     & 1.45 \\
				$\rm ^{104}Zr$ &       &      & 0.037 & 1.01 & 0.002 & 0.22 \\
				$\rm ^{102}Mo$ &       &      & 0.025 & 0.08 & 0.002 & 0.46 \\
				$\rm ^{104}Mo$ &       &      & 0.031 & 1.08 & 0.001 & 2.83 \\
				$\rm ^{106}Mo$ &       &      &       &      & 0.001 & 3.47 \\
				$\rm ^{108}Mo$ &       &      &       &      & 0     & 0.67 \\
				$\rm ^{108}Ru$ &       &      &       &      & 0     & 1.98 \\
				$\rm ^{110}Ru$ &       &      &       &      & 0.011 & 3.62 \\
				$\rm ^{112}Ru$ &       &      &       &      & 0.014 & 0.94 \\
				$\rm ^{112}Pd$ &       &      &       &      & 0.014 & 0.75 \\
				$\rm ^{114}Pd$ &       &      &       &      & 0.035 & 1.82 \\
				$\rm ^{116}Pd$ &       &      &       &      & 0.038 & 0.82 \\
				$\rm ^{130}Sn$ & 0.031 & 0.84 & 0.028 & 1.65 & 0.050 & 0.36 \\
				$\rm ^{132}Sn$ & 0.023 & 1.54 & 0.027 & 1.88 & 0.037 & 0.14 \\
				$\rm ^{134}Sn$ &       &      & 0.026 & 0.18 &       &      \\
				$\rm ^{132}Te$ & 0.008 & 0.35 & 0.029 & 0.47 &       &      \\
				$\rm ^{134}Te$ & 0.023 & 3.11 & 0.024 & 3.95 & 0.059 & 2.35 \\
				$\rm ^{136}Te$ & 0.006 & 3.44 & 0.013 & 3.53 & 0.050 & 0.91 \\
				$\rm ^{138}Te$ & 0.005 & 0.76 & 0.007 & 0.55 & 0.047 & 3.63 \\
				$\rm ^{138}Xe$ & 0.006 & 2.08 & 0.007 & 2.04 & 0.026 & 2.55 \\
				$\rm ^{140}Xe$ & 0.011 & 5.73 & 0.004 & 4.04 & 0.023 & 0.37 \\
				$\rm ^{142}Xe$ & 0.014 & 2.25 & 0.009 & 1.53 & 0.019 & 2.70 \\
				$\rm ^{142}Ba$ & 0.014 & 0.64 & 0.011 & 0.69 & 0.001 & 3.37 \\
				$\rm ^{144}Ba$ & 0.035 & 4.49 & 0.012 & 2.46 & 0     & 0.98 \\
				$\rm ^{146}Ba$ & 0.040 & 2.76 & 0.014 & 1.98 &       &      \\
				$\rm ^{148}Ba$ &       &      & 0.035 & 0.25 &       &      \\
				$\rm ^{148}Ce$ & 0.056 & 0.56 & 0.036 & 0.75 & 0     & 2.35 \\
				$\rm ^{150}Ce$ & 0.079 & 0.41 & 0.067 & 0.86 & 0.004 & 0.94 \\
				$\rm ^{152}Nd$ &       &      &       &      & 0.009 & 0.83 \\
				$\rm ^{154}Nd$ &       &      &       &      & 0.020 & 0.42 \\
				\hline
				$\tilde{c}_{J_1J_2}$ &
				\multicolumn{2}{c|}{0.034}  &
				\multicolumn{2}{c|}{0.020}  &
				\multicolumn{2}{c}{0.017}
			\end{tabular}
		\end{ruledtabular}
	\end{table}
	
	Now, let us consider the results of estimating the correlation coefficients $\tilde{c}_{J_1 J_2}$ in a different theoretical approach, realized in the framework of the temperature model \texttt{FREYA}~\cite{randrup2009,verbeke2018} proposed by Randrup's group. This statistical model employs the Monte Carlo method to generate macroscopic characteristics of primary FFs, including their mass, charge, and initial velocities following the rupture of a compound nucleus. To incorporate dynamics, a nucleon exchange process is incorporated into the aforementioned procedure, which responds to the change in the spins of the FFs at the scission point by using the six rotational modes: doubly degenerate transverse wriggling and bending, and longitudinal twisting and tilting, as outlined in ~\cite{dossing_I_1985, dossing_II_1985}. After that, the thermalization process of FF by sequential evaporation of neutrons and gamma rays is considered, thereby generating large samples of complete fission events. However, upon investigating the mobility coefficients of the previously mentioned rotational modes, it was observed~\cite{randrup2009} that they differ significantly in magnitude. The relaxation times for the wriggling mode are markedly brief, whereas the tilting mode is excited at a relatively slow rate. Consequently, in contrast to the statistical equilibrium model for the FF spins proposed by~\cite{moretto1980}, which is responsible for the nucleon exchange, the latest \texttt{FREYA} approach implements a simplified mechanism assuming only complete relaxation of the transverse modes, as described in~\cite{vogt2013,randrup2014}.
	
	Therefore, in order to describe the spin distributions, the \texttt{FREYA} considers only four modes of transverse wriggling and bending vibrations. Accordingly, the amplitude of the $s_k$ mode of oscillation is determined by a density function of the following form:
	\begin{equation}\label{eq:14}
		P( s_k )\sim \exp\left[ -\frac{s_k^2}{2I_kT_s} \right],
	\end{equation}
	where $I_k$ is the moment of inertia for the considered oscillation mode~\cite{randrup2021,vogt2021,randrup_dossing2022}, and the moments of inertia of wriggling and bending vibrations~\cite{verbeke2018} are related to the FF moments of inertia as
	\begin{equation}\label{eq:15}
		I_{\rm w} =\frac{(I_1 + I_2) I}{I_0}; \qquad I_b = \frac{I_1 I_2}{I_1 + I_2}.
	\end{equation}
	In the context of the distribution described in~\cref{eq:14}, the temperature $T_s$ accounts for the fact that a portion of the excitation energy is transferred to the deformation energy of the fission pre-fragments. Consequently, the energy released during the decay of the fission system is somewhat reduced.
	
	In addition, the relation between the spins of the fragments and the spins of the wriggling and bending vibrations, taking into account the moments of inertia, was constructed in~\cite{randrup2022}, which is of the form:
	\begin{equation}\label{eq:16}
		J_i = \frac{I_i}{I}{J_{\rm w}} + (-1)^{i+1} J_b,
	\end{equation}
	and their variance was defined as
	\begin{equation}\label{eq:17}
		\sigma_i^2 = 2(1- I_i / I ) I_i T_s.
	\end{equation}
	The authors demonstrate that by modeling fission pre-fragments near the scission point using realistic values of the moments of inertia suggested by the "at hot" approximation, it is possible to obtain results that align with experimental observations. This approach is outlined in~\cite{randrup2021} and further developed in~\cite{randrup2022}. The approximation posits that for magic nuclei, the moment of inertia is taken to be $0.2 I_{rigid}$, while for other nuclei, it assumes values of $0.5 I_{rigid}$. As previously noted in~\cite{randrup2021}, this choice of moments of inertia naturally gives rise to a sawtooth-like behaviour of the average spin of the fragments, which provides an explanation for the observation that a light fragment can exhibit a larger spin than a heavy one. Furthermore, the authors contend that the spins of the FFs are not independent due to the angular momentum conservation law. As has been previously established~\cite{kadmensky2017,kadmensky_yad2024}, the spin contributions of the two fragments for wriggling vibrations are parallel, while the contributions for bending vibrations are antiparallel.
	
	Then, using the statistical approach developed in~\cite{randrup2022}, the generalized correlation coefficient for individual FF spins is
	\begin{equation}\label{eq:18}
		c\left(S_L, S_H \right)=\frac{\sigma \left(S_L, S_H \right)}{\sigma \left(S_L\right)\sigma \left( S_H \right)},
	\end{equation}
	or
	\begin{equation}\label{eq:19}
		\begin{aligned}
			&c\left(S_L, S_H \right) = \frac{\left\langle S_L \cdot S_H \right\rangle - \langle S_L \rangle \langle S_H \rangle}{\sigma_L \sigma_H} = \\ &\qquad =-\sqrt{\frac{I_L I_H}{(I_R + I_L)( I_R + I_H)}},
		\end{aligned}
	\end{equation}
	where $I_R$ represents the moment of inertia of the relative motion of the fissile nucleus, and $I_L$ and $I_H$ are analogous for the light and heavy fragment, respectively.
	
	Nevertheless, the value yielded by the formula in \cref{eq:19} is typically quite modest, given that $I_R$ is an order of magnitude larger than that of the individual fragments, $I_L$ and $I_H$. Accordingly, the authors posit that the resulting spins will be relatively independent, though strongly related to the spins of the wriggling and bending vibrations.
	
	This assumption is corroborated by simulations conducted with the \texttt{FREYA} code, the results of which are presented in~\Cref{tab:t3}. Here, the mean spin values and correlation coefficients for the four actinide nuclei that have been studied most extensively are tabulated. The generalized correlation coefficients are close to zero, indicating that the primary spin values are largely uncorrelated. This finding is in agreement with experimental data presented in~\cite{wilson2021}.
	
	\begin{table}
		\caption{\label{tab:t3}The mean values of the primary FF spins $\langle S_L \rangle$ and $\langle S_H \rangle$, and the corresponding correlation coefficients $c (S_L, S_H)$ for $\rm ^{235}U (n,f)$, $\rm ^{238}U ( n,f )$, $\rm ^{239}Pu (n,f)$, and $\rm {}^{252}Cf(sf)$ obtained in~\cite{randrup2021}}
		\begin{ruledtabular}
			\begin{tabular}{cccccc}
				& $\rm {}^{235}U \left( n,f \right)$ & $\rm {}^{238}U \left( n,f \right)$ & $\rm {}^{239}Pu \left( n,f \right)$ & $\rm {}^{252}Cf \left( sf \right)$ \\
				\hline
				$\left\langle S_L \right\rangle $  & 4.27 & 4.43  & 4.58  & 5.08   \\
				$\left\langle S_H \right\rangle $  & 5.66 & 5.80  & 5.93  & 6.33   \\
				$c\left( S_L,S_H \right)$  & 0.002 & 0.002  & 0.001  & 0.001   
			\end{tabular}
		\end{ruledtabular}
	\end{table}
	
	The approach of the Randrup group has shown that, based on correlated transverse bending and wriggling vibrations prior to the rupture of the composite system, the FFs obtain substantial spins which, however, are practically independent after the scission. Although the results of the correlation coefficient analysis in the approach of~\cite{randrup2021} and the one proposed in the present work differ by an order of magnitude in absolute values, they confirm that there is no correlation between the spins of the fission fragments, which agrees well with the experimental data~\cite{wilson2021}.
	
	It is clear from the above that both types of transverse vibrations contribute to the spin distributions of FFs. However, there has been a prolonged debate surrounding the appropriate way to account for the contributions of these observed vibrations, which has been ongoing for over 50 years. The authors of~\cite{rasmussen1969} proposed that only the bending mode be used to describe the pumping of large values of spins and FFs spin distributions for the considered types of fission. This is due to the fact that the energy magnitude of the zero-point bending vibrations is considerably smaller than that of the wriggling. Consequently, in the temperature approach, as the authors of~\cite{rasmussen1969} suggest, the leading contribution will be solely from the bending vibrations. This concept is markedly distinct from current approaches, exemplified by the works~\cite{kadmensky2017, randrup2021, vogt2021, stetcu2021, randrup2022, randrup_dossing2022, kadmensky_yad2024}. Some of these studies~\cite{randrup2021, stetcu2021, randrup_dossing2022}, posit that wriggling vibrations are the primary contributor to the distributions. Indeed, this is related to the correct view of wriggling vibrations, which actually determine the distribution of the relative orbital moments of FF. Nevertheless, it is unclear to what extent this phenomenon dominates the contribution from bending vibrations. This is particularly relevant in the context of the developing approach presented in this study.
	
	An analysis of the angular distributions of the spins can be an additional check of the proposed approach to the above evaluation of the $\tilde{c}_{J_1 J_2}$ correlation coefficients. It will allow us to determine more precisely to what extent the contributions of wriggling and bending vibrations differ, and under what conditions one of the oscillation types dominates. We will do this in the next section of present paper.
	
	\section{Angular correlations between fragment spins}
	In addition to the above, there is currently a great deal of interest in the estimation of the angular correlation of the FF spins, given that there are a number of contradictory results for the angular distributions $P_{12}(\varphi)$ of these SDs. This section will compare the predictions of various theoretical models and analyze the angular distributions obtained from our numerical calculations.
	
	The initial efforts to derive estimates of the angular distributions $P_{12} (\varphi)$ were investigated using the previously mentioned \texttt{FREYA}, with the assumption that the spins of the fragments are perpendicular to the symmetry axis of the fissile system. In the context of this two-dimensional representation of the FF spins, it was demonstrated that there is a negligible correlation between them, both in magnitude and direction. This indicates that the distribution of $P_{12}( \varphi )$ is nearly isotropic. 
	
	The other group of theorists, led by A. Bulgac, employs a distinct approach based on the methods of time-dependent density functional theory, which rejects the assumption of perpendicularity of the FF spins to the symmetry axis of the fissile system. In contrast, the aforementioned group employs a three-dimensional model of the spin distribution, which yields an angular distribution that peaks near the $135^{\circ}$ angle. This result is qualitatively different from that observed by the Randrup group. In more recent work~\cite{scamps2023}, changes were made to the model, resulting in a marked change in the angular distributions. The previously observed maximum disappeared, but two maxima appeared, the larger in the neighborhood of $20^{\circ}$, the smaller at $165^{\circ}$, with a nearly isotropic distribution in between.
	
	Furthermore, we highlight the approach of the University of Tokyo group~\cite{chen2023} within the framework of the antisymmetric molecular dynamics model. In this approach, angular distributions were constructed with a maximum occurring in the region close to $90^{\circ}$. In other words, there are several competing theoretical models in this problem. For this reason, our investigation will extend beyond an analysis of the impact of transverse vibrations. We will also undertake a comparative analysis of the angular distributions obtained within the framework of one of these approaches. To this end, we must integrate the formula~\eqref{eq:8} over the spins of both FFs, thereby obtaining a distribution that depends solely on the angle between the FF spins, given by $w(\varphi)$. 
	
\begin{figure}
		\centering
		\includegraphics[width=\linewidth]{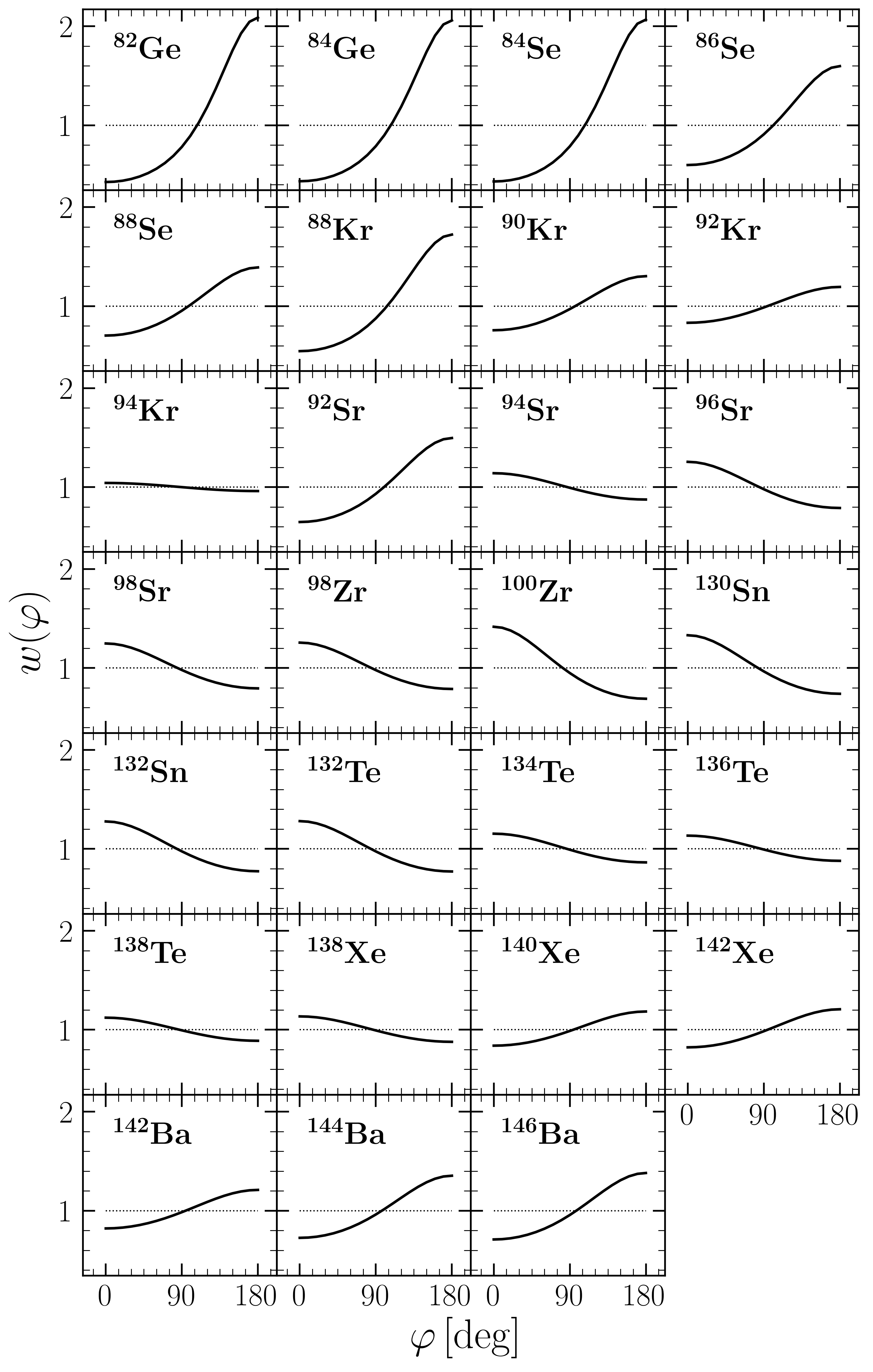}
		\vspace{-2em}
		\caption{The distribution of the opening angle $\varphi$ between the angular momenta of the FFs from $\rm ^{232}Th(n,f)$}\label{fig:f1}
		\vspace{-1em}
	\end{figure}

\begin{figure}
		\centering
		\includegraphics[width=\linewidth]{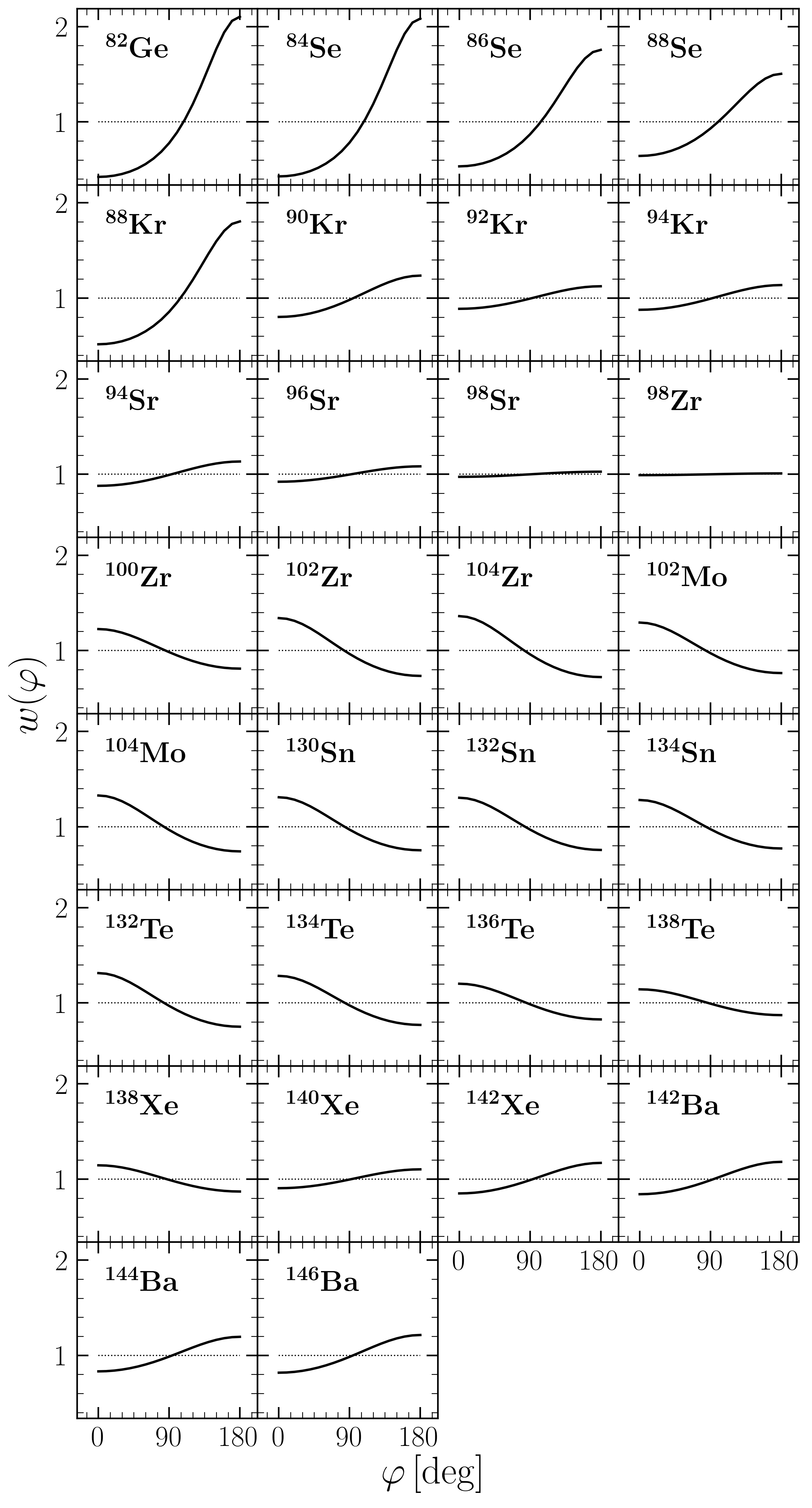}
		\vspace{-2em}
		\caption{The distribution of the opening angle $\varphi$ between the angular momenta of the FFs from $\rm ^{238}U(n,f)$}\label{fig:f2}
		\vspace{-1em}
	\end{figure}

\begin{figure}
		\centering
		\includegraphics[width=1.035\linewidth]{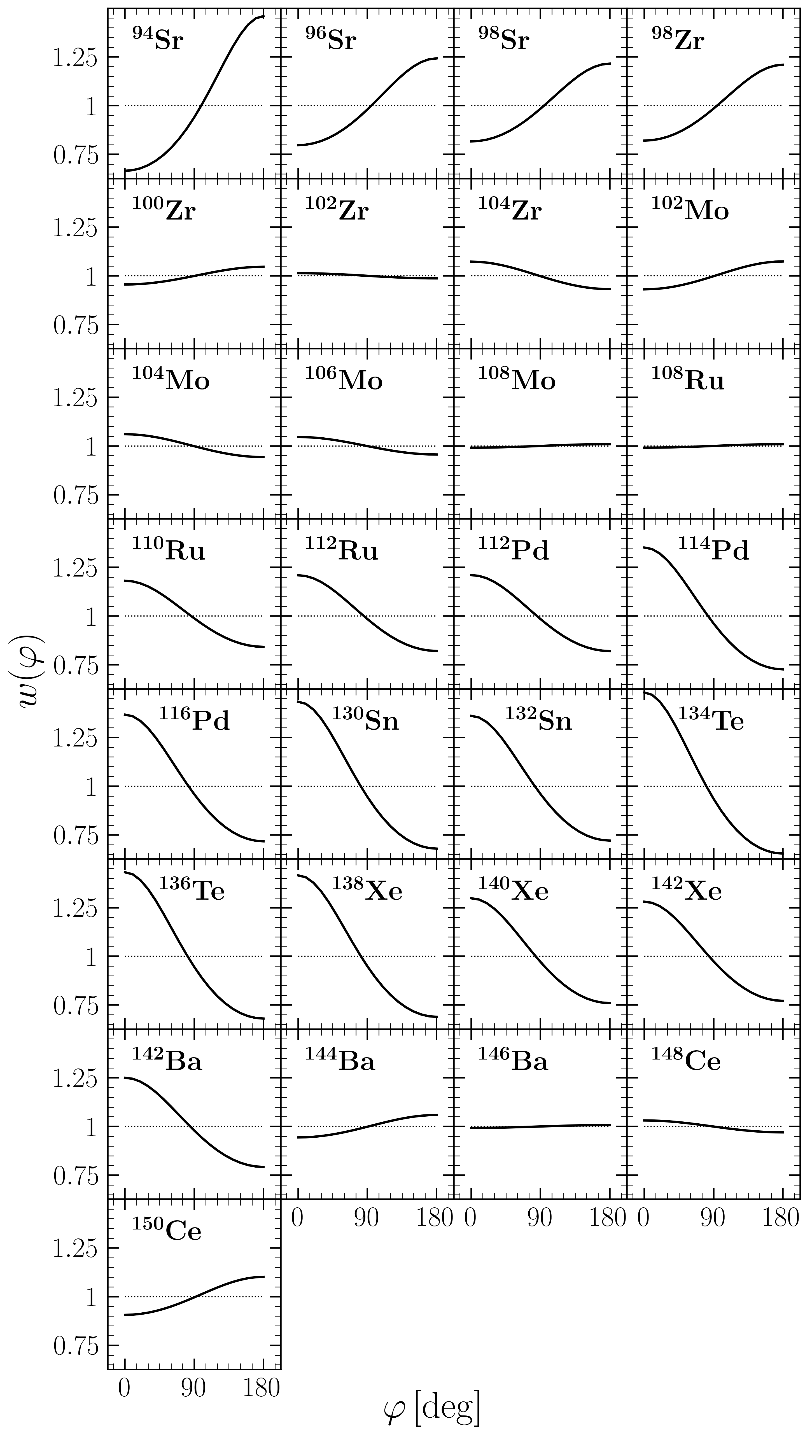}
		\vspace{-2em}
		\caption{The distribution of the opening angle $\varphi$ between the angular momenta of the FFs from $\rm ^{252}Cf(sf)$}\label{fig:f3}
		\vspace{-1em}
	\end{figure}

	It is regretfully noted that an analytical dependence could not be obtained, therefore a numerical calculation was necessary. This procedure was carried out for each pair of FFs as given~\Cref{tab:t1} and~\Cref{tab:t2}. The angular spin distributions of these fragments, calculated using the aforementioned methodology, are presented in~\cref{fig:f1,fig:f2,fig:f3}, respectively, for the reactions  $\rm ^{232}Th(n,f)$, $\rm ^{238}U(n,f)$, and $\rm ^{252}Cf(sf)$.
	
\begin{figure}
		\centering
		\includegraphics[width=\linewidth]{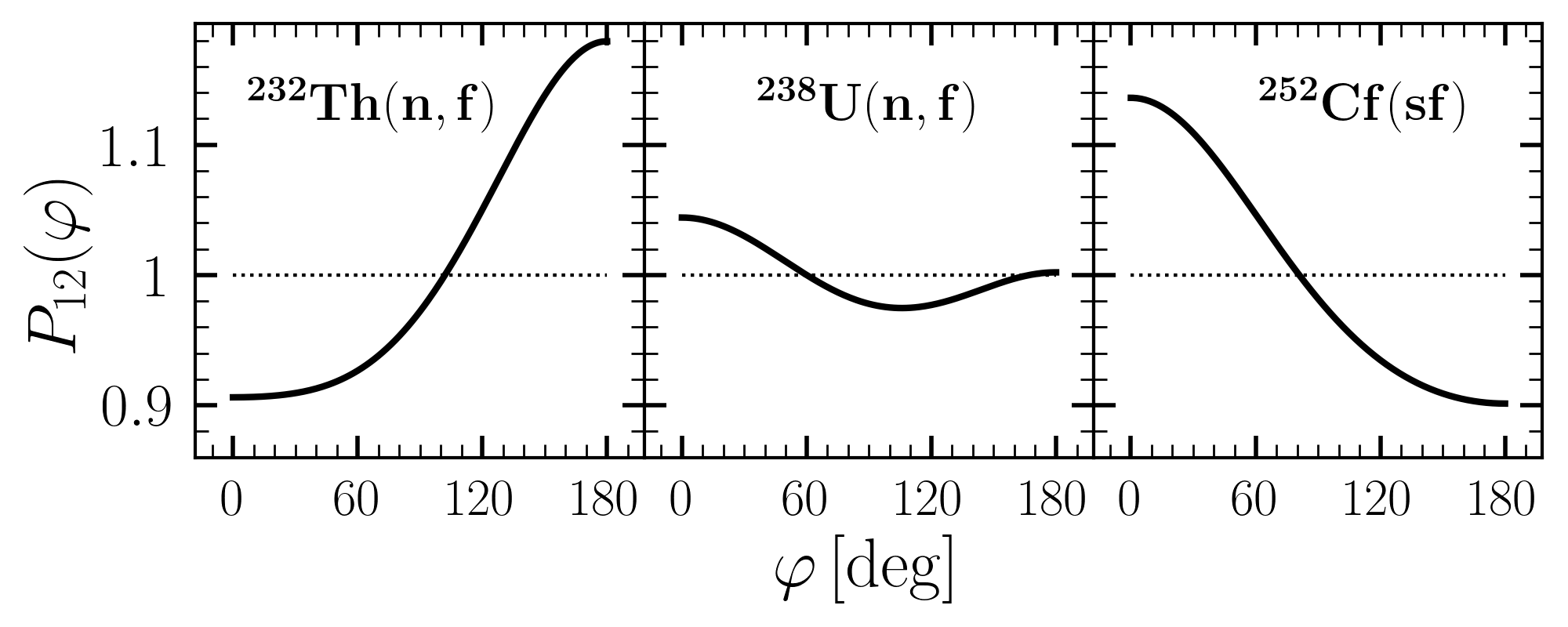}
		\vspace{-2em}
		\caption{Total angular distributions for the three investigated reactions}\label{fig:f4}
		\vspace{-1em}
	\end{figure}

	Concurrently, employing the identical calculation methodology utilized to ascertain the generalized correlation coefficient~\cref{eq:13}, we can derive the integral angular distributions of spins for these nuclei, as illustrated in~\cref{fig:f4}. This allows us to undertake a detailed analysis.
	
	The angular distributions of the spins for the reactions $\rm ^{232}Th(n,f)$, $\rm ^{238}U(n,f)$, and $\rm ^{252}Cf(sf)$, as illustrated in~\cref{fig:f4} reveal distinct patterns that reflect the interplay between wriggling and bending vibrational modes. These modes are strongly influenced by the mass and charge asymmetry of the fission fragments, which vary significantly across the three reactions. To quantify the relative contributions of these modes, we introduce the coefficient $\frac{w(180^{\circ}) - w(0^{\circ})}{w(180^{\circ}) + w(0^{\circ})}$, which measures the preference for antiparallel (bending-dominated) versus parallel (wriggling-dominated) spin configurations.
	
	For $\rm ^{232}Th(n,f)$, where bending vibrations dominate, the fragments typically exhibit significant mass and charge asymmetry. In such cases, the stiffness coefficients for bending modes are reduced, making these modes more favorable. This results in a higher probability of spins being oriented in opposite directions, as bending vibrations inherently lead to antiparallel spin configurations. The coefficient for this reaction, calculated as $13.5\%$, confirms the predominance of bending modes. Conversely, for $\rm ^{252}Cf(sf)$, where wriggling vibrations are more prevalent, the fragments are characterized by greater mass and charge symmetry. Here, the stiffness coefficients for wriggling modes are enhanced, leading to a higher probability of spins being co-directional. The coefficient for this reaction, calculated as $12\%$, underscores the preference for parallel spin configurations, consistent with the nature of wriggling vibrations. The reaction $\rm ^{238}U(n,f)$ represents an intermediate scenario, where the contributions of wriggling and bending vibrations are nearly balanced. The coefficient of $2\%$ indicates that the probability of co-directional and opposite-direction spins is approximately equal, reflecting the moderate level of mass and charge asymmetry of the fragments in this reaction.
	
	The observed behavior can be understood through the lens of mass and charge distributions, which directly influence the stiffness coefficients of the vibrational modes. For asymmetric fragments, such as those in $\rm ^{232}Th(n,f)$, bending modes are favored due to their lower stiffness coefficients, which arise from the complex nuclear and Coulomb potentials associated with large mass and charge asymmetry. In contrast, for more symmetric fragments, such as those in $\rm ^{252}Cf(sf)$, the stiffness coefficients for wriggling modes increase, making these modes more dominant. The intermediate case of $\rm ^{238}U(n,f)$ reflects a balance between these two extremes, where the moderate asymmetry of the fragments results in nearly equal contributions from both modes.
	
	A particularly intriguing aspect of these findings is the non-linear transition from thorium to uranium and, ultimately, to californium. Despite the relatively minor differences in mass distributions between $\rm ^{232}Th(n,f)$ and $\rm ^{238}U(n,f)$, the bending mode contribution drops sharply, as evidenced by the rapid decrease in the coefficient from $13.5\%$ to $2\%$. This suggests that even a slight increase in the yield of symmetric fission fragments can significantly alter the balance between wriggling and bending modes. However, the subsequent transition to $\rm ^{252}Cf(sf)$ reveals a more gradual increase in the dominance of wriggling modes, with the coefficient rising to $12\%$. This slower growth in the wriggling mode contribution, compared to the rapid decline in the bending mode contribution, raises important questions about the underlying mechanisms governing this transition. Specifically, it remains unclear why the bending mode's influence diminishes so rapidly between thorium and uranium, while the wriggling mode's dominance grows more slowly toward californium. This non-monotonic behavior highlights the complex relationship between fragment asymmetry and the dominance of vibrational modes, warranting further investigation with a broader range of nuclei to precisely identify the critical thresholds and physical mechanisms driving these transitions.
	
\begin{figure}
		\centering
		\includegraphics[width=.9\linewidth]{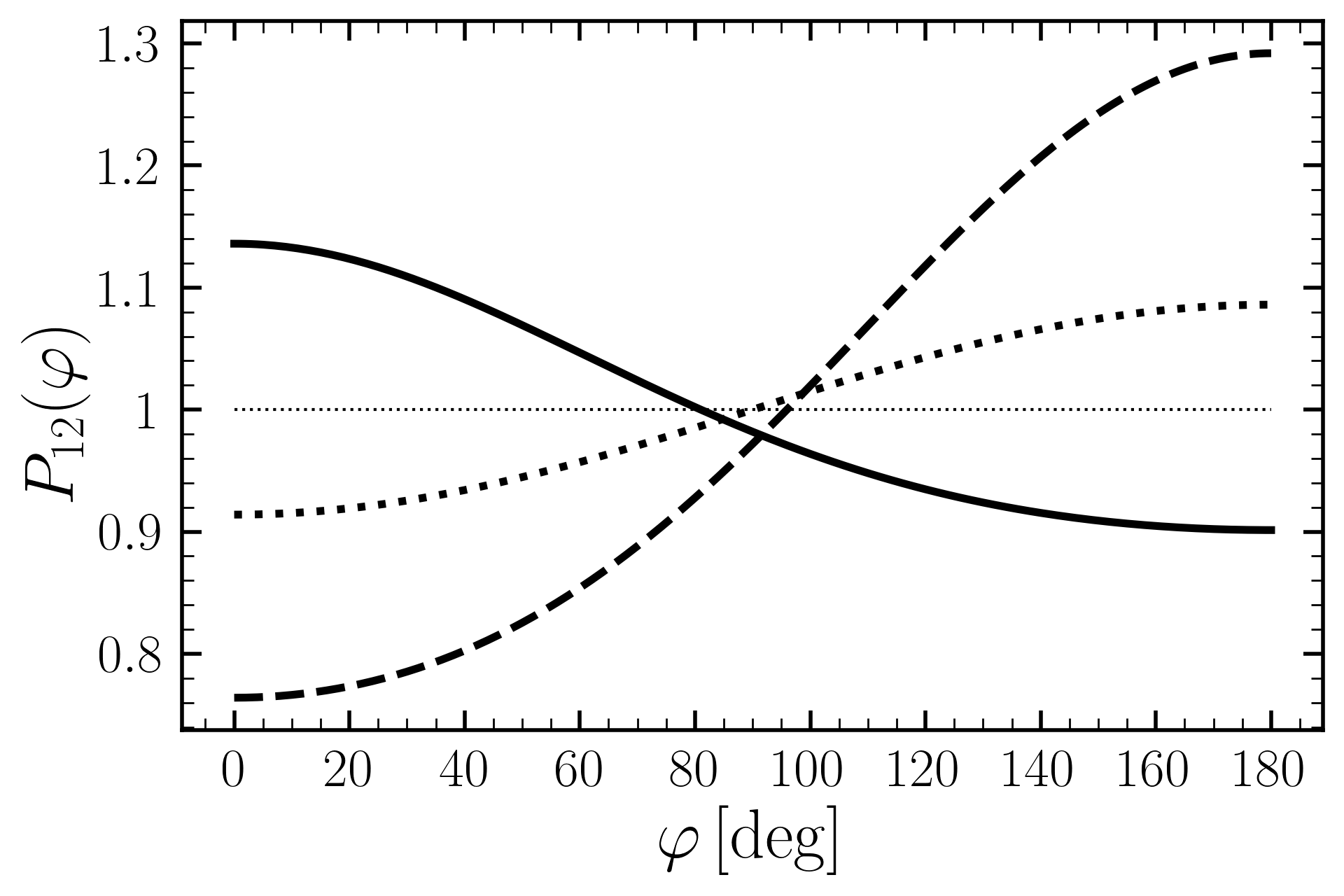}
		\vspace{-1em}
		\caption{Comparison of SD dependence on angle for the response $\rm ^{252} Cf (sf)$. The solid line is the result of the present study, the short and long dashed lines are the first and second limiting cases of the Randrup~\cite{randrup2021,vogt2021} approach}\label{fig:f5}
	\end{figure}

	As our model utilizes a two-dimensional approximation to describe the spin, it is most reasonable to compare the resulting distribution with the findings of the Randrup group, as presented in~\cite{randrup2021, vogt2021}. In the aforementioned works, analogous angular distributions were devised for spontaneous fission of $\rm ^{252}Cf(sf)$ within the temperature approach, encompassing both limiting cases. In the first scenario, the moment of inertia of the relative motion is significantly larger than the sum of the two moments of inertia of the fragments. This results in the conclusion that the distribution of $\bar{P}_{12}(\varphi)$ in the plane perpendicular to the symmetry axis will be isotropic. Furthermore, when considering the connection with the orbital moment of the FF, a slight deviation from this isotropic character is observed. It is accurately approximated by a Fourier expansion of lower orders, that is, by the function $\bar{P}_{12} (\varphi) \approx 1 + \bar{f}_1 \cos\varphi$, where the coefficient of deviation from the isotropic distribution, $\bar{f}_1$, is negligible and equal to $-0.086$. In the alternative scenario, when the moment of inertia of relative motion is comparable to the sum of two moments of inertia of individual fragments, $\tilde{P}_{12} (\varphi)$ there are significant deviations from the isotropic character, due to the larger value of the coefficient $\tilde{f}_1 = - 0.264$. Therefore, for a qualitative description, it becomes necessary to supplement the expansion with a second-order term, $\tilde{P}_{12} (\varphi) \approx 1+ \tilde{f}_1 \cos \varphi + \tilde{f}_2 \cos 2\varphi$, where $\tilde{f}_2 = 0.028$. Both cases considered are presented in~\cref{fig:f5} along with the distribution from~\cref{fig:f4} obtained in this paper.
	
	As illustrated in~\Cref{fig:f5}, wriggling vibrations are predominant, given that the probability of spin co-orientation is approximately $20\%$ greater than the probability of their opposite direction. In contrast, the Randrup approach exhibits the opposite trend, with probabilities for the limiting cases of about $18\%$ and $68\%$ by modulus, respectively.
	
	Therefore, the hypothesis proposed in~\cite{randrup2022}, which suggests that the moment of inertia of relative motion serves as an effective accumulator of orbital momentum, has been validated by the analysis presented in this study. This occurs because the ratio of relative motion moments of inertia to the sum of FF moments of inertia was shown to be in the range of 4-5, which is consistent with the results presented in~\cite{kadmensky_ran2024}. Additionally, a crucial pattern was identified: an increase in the anisotropy of the angular distribution is associated with an rise in the correlation coefficient, suggesting a direct correlation between these characteristics. The physical explanation of this phenomenon is relatively straightforward. It can be attributed to the strong correlation between the spins of the fragments, which reduces the isotropy of their angular distribution. This dependence is clearly evident upon analysis of the partial correlation coefficients and angular distributions presented in~\Cref{tab:t2} and~\cref{fig:f1,fig:f2,fig:f3}, respectively.
	
	The key distinction between our approach and the model proposed by the Randrup group lies in the utilization of the concept of a ``cold'' compound nucleus, as opposed to the temperature-based approach employed in their studies. It is noteworthy that the generalized correlation coefficients obtained in the cited papers~\cite{randrup2021,randrup2022} were an order of magnitude lower than those calculated in the present study. However, the anisotropies of the angular distributions in both studies were found to be comparable. This result raises questions and requires a more detailed theoretical analysis.
	
	In light of the foregoing comparison, we may briefly mention the results obtained using the TDDFT approach, in which both three-dimensional~\cite{bulgac2021, stetcu2021, scamps2023, scamps2024} and two-dimensional~\cite{scamps2024} spin descriptions were applied. Of particular interest for our study are the results of the latter work, in which we compared the angular distributions with data obtained using the \texttt{FREYA} model~\cite{vogt2021}, based on the two-dimensional spin model. The distribution was found to be nearly isotropic in the majority of cases, particularly within the angle range of $40^{\circ}$ to $160^{\circ}$, with a mean probability value of $0.06$. However, the first maximum around $25^{\circ}$ reaches $0.17$, which is approximately three times the mean value. In the three-dimensional model, the distribution is similar, but with a less pronounced first maximum due to the inclusion of twisting and tilting modes.
	
	Meanwhile, the author of the referenced work~\cite{scamps2024} did not devote sufficient attention to the interpretation of the results. The data obtained indicate that wriggling vibrations are the dominant contribution, while the second, less pronounced maximum in the neighborhood of $165^{\circ}$ with a probability of $0.08$ indicates a small contribution of bending vibrations. It is noteworthy that in our model for the nucleus of $\rm ^{252}Cf$, the dominance of wriggling vibrations is only $20\%$, whereas in the aforementioned work, it reaches almost $300\%$. Furthermore, a comparison with the three-dimensional distribution, in which the first maximum corresponding to wriggling vibrations is reduced to $0.12$, corroborates the hypothesis proposed by the Randrup group. This demonstrates that the longitudinal vibrations included in the three-dimensional description do not exert a dominant influence due to the shorter relaxation time in comparison to the transverse vibrations.
	
	Therefore, given the significant discrepancies between the theoretical models of different groups with respect to the angular distributions and correlation coefficients of the FF spins, it becomes crucial to have experimental data that could confirm or refute the theoretical predictions. This could help elucidate the processes of spontaneous and induced low-energy fission and clarify the underlying mechanisms of these phenomena.

	\section{CONCLUSION}
	
	In the present study, we developed a theoretical model~\cite{kadmensky2017,kadmensky_yad2024,lyubashevsky2025}, which was based on the quantum fission theory~\cite{kadmensky2002, kadmensky2003, kadmensky2005, kadmensky2008}. This allowed us to analyze the key characteristics of the SD of the fission products of the following reactions: $\rm ^{232}Th(n,f)$, $\rm ^{238}U(n,f)$, and $\rm ^{252}Cf(sf)$. The model was able to account for the absence of a correlation between the spins of light and heavy FFs observed in the experiment. Importantly, this finding is in reasonable agreement with the results presented in~\cite{wilson2021}. Nevertheless, the interpretation of the experiment presented in the paper, specifically regarding the formation of FF spins following the rupture of the compound nucleus, was not corroborated. In the context of our model, which utilizes zero bending and wriggling vibrations to describe SD, it has been demonstrated that FF spins remain uncorrelated even prior to the fissile nucleus rupture. This finding aligns with the analogous conclusion presented by the theoretical Randrup group in~\cite{randrup2021, vogt2021, randrup_dossing2022}.
	
	Our analysis of the ADs of spins revealed that bending vibrations play a prominent role in the fission reaction of $\rm \rm ^{232}Th(n,f)$, whereas wriggling vibrations are dominant in the reactions of $\rm ^{238}U(n,f)$ and $\rm ^{252}Cf(sf)$. This highlights the importance of considering both types of vibrational modes in the description of fission dynamics. A comparison of the obtained data with the results of other theoretical groups revealed that the proposed approach, based on the ``cold'' nucleus concept, offers advantages in terms of simplicity in describing the ADs and spin correlation coefficients. However, some discrepancies, particularly in the values of the correlation coefficients, necessitate further theoretical analysis and additional verification.
	
	The developed model offers a significant contribution to the understanding of heavy nucleus fission processes, providing new perspectives for the study of the spin characteristics of FFs. The results obtained not only extend the existing theory of fission but also indicate the necessity for additional experimental data to verify theoretical predictions, for example, the direct relation between the correlation coefficient and angular anisotropy, and to clarify the mechanism of spin formation in fission reactions.

	\begin{acknowledgments}
	The authors are grateful to Professor S.G. Kadmensky for his active discussion on the subject of quantum fission theory, particularly for his insights into the role of vibrational modes in spin formation. We also thank Associate Professor L.V. Titova for her valuable comments and clarifications during the preparation of this paper. Special thanks to Dr. Guillaume Scamps for his detailed feedback and clarifications on the model developed by the Bulgac group, which helped us refine our theoretical approach.
	\end{acknowledgments}
	
	\bibliographystyle{apsrev4-2}
	\bibliography{bibliography.bib}

\providecommand{\noopsort}[1]{}\providecommand{\singleletter}[1]{#1}%
\begin{thebibliography}{41}%
\makeatletter
\providecommand \@ifxundefined [1]{%
 \@ifx{#1\undefined}
}%
\providecommand \@ifnum [1]{%
 \ifnum #1\expandafter \@firstoftwo
 \else \expandafter \@secondoftwo
 \fi
}%
\providecommand \@ifx [1]{%
 \ifx #1\expandafter \@firstoftwo
 \else \expandafter \@secondoftwo
 \fi
}%
\providecommand \natexlab [1]{#1}%
\providecommand \enquote  [1]{``#1''}%
\providecommand \bibnamefont  [1]{#1}%
\providecommand \bibfnamefont [1]{#1}%
\providecommand \citenamefont [1]{#1}%
\providecommand \href@noop [0]{\@secondoftwo}%
\providecommand \href [0]{\begingroup \@sanitize@url \@href}%
\providecommand \@href[1]{\@@startlink{#1}\@@href}%
\providecommand \@@href[1]{\endgroup#1\@@endlink}%
\providecommand \@sanitize@url [0]{\catcode `\\12\catcode `\$12\catcode
  `\&12\catcode `\#12\catcode `\^12\catcode `\_12\catcode `\%12\relax}%
\providecommand \@@startlink[1]{}%
\providecommand \@@endlink[0]{}%
\providecommand \url  [0]{\begingroup\@sanitize@url \@url }%
\providecommand \@url [1]{\endgroup\@href {#1}{\urlprefix }}%
\providecommand \urlprefix  [0]{URL }%
\providecommand \Eprint [0]{\href }%
\providecommand \doibase [0]{https://doi.org/}%
\providecommand \selectlanguage [0]{\@gobble}%
\providecommand \bibinfo  [0]{\@secondoftwo}%
\providecommand \bibfield  [0]{\@secondoftwo}%
\providecommand \translation [1]{[#1]}%
\providecommand \BibitemOpen [0]{}%
\providecommand \bibitemStop [0]{}%
\providecommand \bibitemNoStop [0]{.\EOS\space}%
\providecommand \EOS [0]{\spacefactor3000\relax}%
\providecommand \BibitemShut  [1]{\csname bibitem#1\endcsname}%
\let\auto@bib@innerbib\@empty
\bibitem [{\citenamefont {Wilson}\ \emph {et~al.}(2021)\citenamefont {Wilson}
  \emph {et~al.}}]{wilson2021}%
  \BibitemOpen
  \bibfield  {author} {\bibinfo {author} {\bibfnamefont {J.~N.}\ \bibnamefont
  {Wilson}} \emph {et~al.},\ }\href
  {https://doi.org/10.1038/s41586-021-03304-w} {\bibfield  {journal} {\bibinfo
  {journal} {Nature}\ }\textbf {\bibinfo {volume} {590}},\ \bibinfo {pages}
  {566} (\bibinfo {year} {2021})}\BibitemShut {NoStop}%
\bibitem [{\citenamefont {Randrup}\ and\ \citenamefont
  {Vogt}(2021)}]{randrup2021}%
  \BibitemOpen
  \bibfield  {author} {\bibinfo {author} {\bibfnamefont {J.}~\bibnamefont
  {Randrup}}\ and\ \bibinfo {author} {\bibfnamefont {R.}~\bibnamefont {Vogt}},\
  }\href {https://doi.org/10.1103/PhysRevLett.127.062502} {\bibfield  {journal}
  {\bibinfo  {journal} {Phys. Rev. Lett.}\ }\textbf {\bibinfo {volume} {127}},\
  \bibinfo {pages} {062502} (\bibinfo {year} {2021})}\BibitemShut {NoStop}%
\bibitem [{\citenamefont {Bulgac}\ \emph {et~al.}(2021)\citenamefont {Bulgac}
  \emph {et~al.}}]{bulgac2021}%
  \BibitemOpen
  \bibfield  {author} {\bibinfo {author} {\bibfnamefont {A.}~\bibnamefont
  {Bulgac}} \emph {et~al.},\ }\href
  {https://doi.org/10.1103/PhysRevLett.126.142502} {\bibfield  {journal}
  {\bibinfo  {journal} {Phys. Rev. Lett.}\ }\textbf {\bibinfo {volume} {126}},\
  \bibinfo {pages} {142502} (\bibinfo {year} {2021})}\BibitemShut {NoStop}%
\bibitem [{\citenamefont {Rasmussen}\ \emph {et~al.}(1969)\citenamefont
  {Rasmussen}, \citenamefont {Nörenberg},\ and\ \citenamefont
  {Mang}}]{rasmussen1969}%
  \BibitemOpen
  \bibfield  {author} {\bibinfo {author} {\bibfnamefont {J.~O.}\ \bibnamefont
  {Rasmussen}}, \bibinfo {author} {\bibfnamefont {W.}~\bibnamefont
  {Nörenberg}},\ and\ \bibinfo {author} {\bibfnamefont {H.~J.}\ \bibnamefont
  {Mang}},\ }\href {https://doi.org/10.1016/0375-9474(69)90066-9} {\bibfield
  {journal} {\bibinfo  {journal} {Nucl. Phys. A}\ }\textbf {\bibinfo {volume}
  {136}},\ \bibinfo {pages} {465} (\bibinfo {year} {1969})}\BibitemShut
  {NoStop}%
\bibitem [{\citenamefont {Nix}\ and\ \citenamefont
  {Swiatecki}(1965)}]{nix1965}%
  \BibitemOpen
  \bibfield  {author} {\bibinfo {author} {\bibfnamefont {J.~R.}\ \bibnamefont
  {Nix}}\ and\ \bibinfo {author} {\bibfnamefont {W.~J.}\ \bibnamefont
  {Swiatecki}},\ }\href {https://doi.org/10.1016/0029-5582(65)90748-5}
  {\bibfield  {journal} {\bibinfo  {journal} {Nucl. Phys. A}\ }\textbf
  {\bibinfo {volume} {71}},\ \bibinfo {pages} {1} (\bibinfo {year}
  {1965})}\BibitemShut {NoStop}%
\bibitem [{\citenamefont {Vogt}\ and\ \citenamefont
  {Randrup}(2021)}]{vogt2021}%
  \BibitemOpen
  \bibfield  {author} {\bibinfo {author} {\bibfnamefont {R.}~\bibnamefont
  {Vogt}}\ and\ \bibinfo {author} {\bibfnamefont {J.}~\bibnamefont {Randrup}},\
  }\href {https://doi.org/10.1103/PhysRevC.103.014610} {\bibfield  {journal}
  {\bibinfo  {journal} {Phys. Rev. C}\ }\textbf {\bibinfo {volume} {103}},\
  \bibinfo {pages} {014610} (\bibinfo {year} {2021})}\BibitemShut {NoStop}%
\bibitem [{\citenamefont {Stetcu}\ \emph {et~al.}(2021)\citenamefont {Stetcu}
  \emph {et~al.}}]{stetcu2021}%
  \BibitemOpen
  \bibfield  {author} {\bibinfo {author} {\bibfnamefont {I.}~\bibnamefont
  {Stetcu}} \emph {et~al.},\ }\href
  {https://doi.org/10.1103/PhysRevLett.127.222502} {\bibfield  {journal}
  {\bibinfo  {journal} {Phys. Rev. Lett.}\ }\textbf {\bibinfo {volume} {127}},\
  \bibinfo {pages} {222502} (\bibinfo {year} {2021})}\BibitemShut {NoStop}%
\bibitem [{\citenamefont {Kadmensky}(2002)}]{kadmensky2002}%
  \BibitemOpen
  \bibfield  {author} {\bibinfo {author} {\bibfnamefont {S.~G.}\ \bibnamefont
  {Kadmensky}},\ }\href {https://doi.org/10.1134/1.1515841} {\bibfield
  {journal} {\bibinfo  {journal} {Phys. At. Nucl.}\ }\textbf {\bibinfo {volume}
  {65}},\ \bibinfo {pages} {1785} (\bibinfo {year} {2002})}\BibitemShut
  {NoStop}%
\bibitem [{\citenamefont {Kadmensky}\ and\ \citenamefont
  {Rodionova}(2003)}]{kadmensky2003}%
  \BibitemOpen
  \bibfield  {author} {\bibinfo {author} {\bibfnamefont {S.~G.}\ \bibnamefont
  {Kadmensky}}\ and\ \bibinfo {author} {\bibfnamefont {L.~V.}\ \bibnamefont
  {Rodionova}},\ }\href {https://doi.org/10.1134/1.1592575} {\bibfield
  {journal} {\bibinfo  {journal} {Phys. At. Nucl.}\ }\textbf {\bibinfo {volume}
  {66}},\ \bibinfo {pages} {1219} (\bibinfo {year} {2003})}\BibitemShut
  {NoStop}%
\bibitem [{\citenamefont {Kadmensky}(2005)}]{kadmensky2005}%
  \BibitemOpen
  \bibfield  {author} {\bibinfo {author} {\bibfnamefont {S.~G.}\ \bibnamefont
  {Kadmensky}},\ }\href {https://doi.org/10.1134/1.2149077} {\bibfield
  {journal} {\bibinfo  {journal} {Phys. At. Nucl.}\ }\textbf {\bibinfo {volume}
  {68}},\ \bibinfo {pages} {1968} (\bibinfo {year} {2005})}\BibitemShut
  {NoStop}%
\bibitem [{\citenamefont {Kadmensky}(2008)}]{kadmensky2008}%
  \BibitemOpen
  \bibfield  {author} {\bibinfo {author} {\bibfnamefont {S.~G.}\ \bibnamefont
  {Kadmensky}},\ }\href {https://doi.org/10.1134/S1063778808070107} {\bibfield
  {journal} {\bibinfo  {journal} {Phys. At. Nucl.}\ }\textbf {\bibinfo {volume}
  {71}},\ \bibinfo {pages} {1226} (\bibinfo {year} {2008})}\BibitemShut
  {NoStop}%
\bibitem [{\citenamefont {Kadmensky}\ and\ \citenamefont
  {Kadmensky}(2009)}]{kadmensky2009}%
  \BibitemOpen
  \bibfield  {author} {\bibinfo {author} {\bibfnamefont {S.~G.}\ \bibnamefont
  {Kadmensky}}\ and\ \bibinfo {author} {\bibfnamefont {S.~S.}\ \bibnamefont
  {Kadmensky}},\ }\href {https://doi.org/10.3103/S1062873809020130} {\bibfield
  {journal} {\bibinfo  {journal} {Bull. Russ. Acad. Sci. Phys.}\ }\textbf
  {\bibinfo {volume} {73}},\ \bibinfo {pages} {193} (\bibinfo {year}
  {2009})}\BibitemShut {NoStop}%
\bibitem [{\citenamefont {Kadmensky}\ \emph
  {et~al.}(2024{\natexlab{a}})\citenamefont {Kadmensky} \emph
  {et~al.}}]{kadmensky_yad2024}%
  \BibitemOpen
  \bibfield  {author} {\bibinfo {author} {\bibfnamefont {S.~G.}\ \bibnamefont
  {Kadmensky}} \emph {et~al.},\ }\href
  {https://doi.org/10.1134/S1063778824600155} {\bibfield  {journal} {\bibinfo
  {journal} {Phys. At. Nucl.}\ }\textbf {\bibinfo {volume} {87}},\ \bibinfo
  {pages} {359} (\bibinfo {year} {2024}{\natexlab{a}})}\BibitemShut {NoStop}%
\bibitem [{\citenamefont {Moretto}\ and\ \citenamefont
  {Schmitt}(1980)}]{moretto1980}%
  \BibitemOpen
  \bibfield  {author} {\bibinfo {author} {\bibfnamefont {L.~G.}\ \bibnamefont
  {Moretto}}\ and\ \bibinfo {author} {\bibfnamefont {R.~P.}\ \bibnamefont
  {Schmitt}},\ }\href {https://doi.org/10.1103/PhysRevC.21.204} {\bibfield
  {journal} {\bibinfo  {journal} {Phys. Rev. C}\ }\textbf {\bibinfo {volume}
  {21}},\ \bibinfo {pages} {204} (\bibinfo {year} {1980})}\BibitemShut
  {NoStop}%
\bibitem [{\citenamefont {Kadmensky}\ \emph {et~al.}(2017)\citenamefont
  {Kadmensky} \emph {et~al.}}]{kadmensky2017}%
  \BibitemOpen
  \bibfield  {author} {\bibinfo {author} {\bibfnamefont {S.~G.}\ \bibnamefont
  {Kadmensky}} \emph {et~al.},\ }\href
  {https://doi.org/10.1134/S1063778817030082} {\bibfield  {journal} {\bibinfo
  {journal} {Phys. At. Nucl.}\ }\textbf {\bibinfo {volume} {80}},\ \bibinfo
  {pages} {447} (\bibinfo {year} {2017})}\BibitemShut {NoStop}%
\bibitem [{\citenamefont {Bohr}\ and\ \citenamefont
  {Wheeler}(1939)}]{bohr1939}%
  \BibitemOpen
  \bibfield  {author} {\bibinfo {author} {\bibfnamefont {N.}~\bibnamefont
  {Bohr}}\ and\ \bibinfo {author} {\bibfnamefont {J.~A.}\ \bibnamefont
  {Wheeler}},\ }\href@noop {} {\bibfield  {journal} {\bibinfo  {journal}
  {Physical Review}\ }\textbf {\bibinfo {volume} {56}},\ \bibinfo {pages} {426}
  (\bibinfo {year} {1939})}\BibitemShut {NoStop}%
\bibitem [{\citenamefont {Bohr}\ and\ \citenamefont
  {Mottelson}(1998)}]{bohr1998}%
  \BibitemOpen
  \bibfield  {author} {\bibinfo {author} {\bibfnamefont {A.}~\bibnamefont
  {Bohr}}\ and\ \bibinfo {author} {\bibfnamefont {B.}~\bibnamefont
  {Mottelson}},\ }\href {https://books.google.com.au/books?id=NNZQDQAAQBAJ}
  {\emph {\bibinfo {title} {Nuclear Structure (In 2 Volumes)}}}\ (\bibinfo
  {publisher} {World Scientific Publishing Company},\ \bibinfo {year}
  {1998})\BibitemShut {NoStop}%
\bibitem [{\citenamefont {Danilyan}(2019)}]{danilyan2019}%
  \BibitemOpen
  \bibfield  {author} {\bibinfo {author} {\bibfnamefont {G.~V.}\ \bibnamefont
  {Danilyan}},\ }\href {https://doi.org/10.1134/S1063778819030050} {\bibfield
  {journal} {\bibinfo  {journal} {Phys. At. Nucl.}\ }\textbf {\bibinfo {volume}
  {82}},\ \bibinfo {pages} {250} (\bibinfo {year} {2019})}\BibitemShut
  {NoStop}%
\bibitem [{\citenamefont {Gagarski}\ \emph {et~al.}(2016)\citenamefont
  {Gagarski}, \citenamefont {G{\"o}nnenwein}, \citenamefont {Guseva},
  \citenamefont {Jesinger}, \citenamefont {Kopatch}, \citenamefont {Kuzmina},
  \citenamefont {Lelievre-Berna}, \citenamefont {Mutterer}, \citenamefont
  {Nesvizhevsky}, \citenamefont {Petrov} \emph {et~al.}}]{gagarski2016}%
  \BibitemOpen
  \bibfield  {author} {\bibinfo {author} {\bibfnamefont {A.}~\bibnamefont
  {Gagarski}}, \bibinfo {author} {\bibfnamefont {F.}~\bibnamefont
  {G{\"o}nnenwein}}, \bibinfo {author} {\bibfnamefont {I.}~\bibnamefont
  {Guseva}}, \bibinfo {author} {\bibfnamefont {P.}~\bibnamefont {Jesinger}},
  \bibinfo {author} {\bibfnamefont {Y.}~\bibnamefont {Kopatch}}, \bibinfo
  {author} {\bibfnamefont {T.}~\bibnamefont {Kuzmina}}, \bibinfo {author}
  {\bibfnamefont {E.}~\bibnamefont {Lelievre-Berna}}, \bibinfo {author}
  {\bibfnamefont {M.}~\bibnamefont {Mutterer}}, \bibinfo {author}
  {\bibfnamefont {V.}~\bibnamefont {Nesvizhevsky}}, \bibinfo {author}
  {\bibfnamefont {G.}~\bibnamefont {Petrov}}, \emph {et~al.},\ }\href@noop {}
  {\bibfield  {journal} {\bibinfo  {journal} {Physical Review C}\ }\textbf
  {\bibinfo {volume} {93}},\ \bibinfo {pages} {054619} (\bibinfo {year}
  {2016})}\BibitemShut {NoStop}%
\bibitem [{\citenamefont {Danilyan}\ \emph {et~al.}(2009)\citenamefont
  {Danilyan} \emph {et~al.}}]{danilyan2009}%
  \BibitemOpen
  \bibfield  {author} {\bibinfo {author} {\bibfnamefont {G.~V.}\ \bibnamefont
  {Danilyan}} \emph {et~al.},\ }\href
  {https://doi.org/10.1016/j.physletb.2009.06.068} {\bibfield  {journal}
  {\bibinfo  {journal} {Phys. Lett. B}\ }\textbf {\bibinfo {volume} {679}},\
  \bibinfo {pages} {25} (\bibinfo {year} {2009})}\BibitemShut {NoStop}%
\bibitem [{\citenamefont {Danilyan}\ \emph {et~al.}(2010)\citenamefont
  {Danilyan} \emph {et~al.}}]{danilyan2010}%
  \BibitemOpen
  \bibfield  {author} {\bibinfo {author} {\bibfnamefont {G.~V.}\ \bibnamefont
  {Danilyan}} \emph {et~al.},\ }\href
  {https://doi.org/10.1134/S1063778810070045} {\bibfield  {journal} {\bibinfo
  {journal} {Phys. At. Nucl.}\ }\textbf {\bibinfo {volume} {73}},\ \bibinfo
  {pages} {1116} (\bibinfo {year} {2010})}\BibitemShut {NoStop}%
\bibitem [{\citenamefont {Valsky}\ \emph {et~al.}(2010)\citenamefont {Valsky}
  \emph {et~al.}}]{valsky2010}%
  \BibitemOpen
  \bibfield  {author} {\bibinfo {author} {\bibfnamefont {G.~V.}\ \bibnamefont
  {Valsky}} \emph {et~al.},\ }\href {https://doi.org/10.3103/S1062873810060080}
  {\bibfield  {journal} {\bibinfo  {journal} {Bull. Russ. Acad. Sci. Phys.}\
  }\textbf {\bibinfo {volume} {74}},\ \bibinfo {pages} {767} (\bibinfo {year}
  {2010})}\BibitemShut {NoStop}%
\bibitem [{\citenamefont {Gagarskii}\ \emph {et~al.}(2011)\citenamefont
  {Gagarskii} \emph {et~al.}}]{gagarskii2011}%
  \BibitemOpen
  \bibfield  {author} {\bibinfo {author} {\bibfnamefont {A.~M.}\ \bibnamefont
  {Gagarskii}} \emph {et~al.},\ }\href
  {https://doi.org/10.1134/S1063774511070133} {\bibfield  {journal} {\bibinfo
  {journal} {Crystallogr. Rep.}\ }\textbf {\bibinfo {volume} {56}},\ \bibinfo
  {pages} {1238} (\bibinfo {year} {2011})}\BibitemShut {NoStop}%
\bibitem [{\citenamefont {Ericson}\ and\ \citenamefont
  {Strutinski}(1958)}]{ericson1958}%
  \BibitemOpen
  \bibfield  {author} {\bibinfo {author} {\bibfnamefont {T.}~\bibnamefont
  {Ericson}}\ and\ \bibinfo {author} {\bibfnamefont {V.}~\bibnamefont
  {Strutinski}},\ }\href {https://doi.org/10.1016/0029-5582(58)90156-1}
  {\bibfield  {journal} {\bibinfo  {journal} {Nucl. Phys.}\ }\textbf {\bibinfo
  {volume} {8}},\ \bibinfo {pages} {284} (\bibinfo {year} {1958})}\BibitemShut
  {NoStop}%
\bibitem [{\citenamefont {Wilhelmy}\ \emph {et~al.}(1972)\citenamefont
  {Wilhelmy} \emph {et~al.}}]{wilhelmy1972}%
  \BibitemOpen
  \bibfield  {author} {\bibinfo {author} {\bibfnamefont {J.~B.}\ \bibnamefont
  {Wilhelmy}} \emph {et~al.},\ }\href {https://doi.org/10.1103/PhysRevC.5.2041}
  {\bibfield  {journal} {\bibinfo  {journal} {Phys. Rev. C}\ }\textbf {\bibinfo
  {volume} {5}},\ \bibinfo {pages} {2041} (\bibinfo {year} {1972})}\BibitemShut
  {NoStop}%
\bibitem [{\citenamefont {Wolf}\ and\ \citenamefont
  {Cheifetz}(1976)}]{wolf1976}%
  \BibitemOpen
  \bibfield  {author} {\bibinfo {author} {\bibfnamefont {A.}~\bibnamefont
  {Wolf}}\ and\ \bibinfo {author} {\bibfnamefont {E.}~\bibnamefont
  {Cheifetz}},\ }\href {https://doi.org/10.1103/PhysRevC.13.1952} {\bibfield
  {journal} {\bibinfo  {journal} {Phys. Rev. C}\ }\textbf {\bibinfo {volume}
  {13}},\ \bibinfo {pages} {1952} (\bibinfo {year} {1976})}\BibitemShut
  {NoStop}%
\bibitem [{\citenamefont {Randrup}\ \emph {et~al.}(2022)\citenamefont
  {Randrup}, \citenamefont {Dössing},\ and\ \citenamefont
  {Vogt}}]{randrup_dossing2022}%
  \BibitemOpen
  \bibfield  {author} {\bibinfo {author} {\bibfnamefont {J.}~\bibnamefont
  {Randrup}}, \bibinfo {author} {\bibfnamefont {T.}~\bibnamefont {Dössing}},\
  and\ \bibinfo {author} {\bibfnamefont {R.}~\bibnamefont {Vogt}},\ }\href
  {https://doi.org/10.1103/PhysRevC.106.014609} {\bibfield  {journal} {\bibinfo
   {journal} {Phys. Rev. C}\ }\textbf {\bibinfo {volume} {106}},\ \bibinfo
  {pages} {014609} (\bibinfo {year} {2022})}\BibitemShut {NoStop}%
\bibitem [{\citenamefont {Shneidman}\ \emph {et~al.}(2002)\citenamefont
  {Shneidman} \emph {et~al.}}]{shneidman2002}%
  \BibitemOpen
  \bibfield  {author} {\bibinfo {author} {\bibfnamefont {T.~M.}\ \bibnamefont
  {Shneidman}} \emph {et~al.},\ }\href
  {https://doi.org/10.1103/PhysRevC.65.064302} {\bibfield  {journal} {\bibinfo
  {journal} {Phys. Rev. C}\ }\textbf {\bibinfo {volume} {65}},\ \bibinfo
  {pages} {064302} (\bibinfo {year} {2002})}\BibitemShut {NoStop}%
\bibitem [{\citenamefont {Migdal}(1960)}]{migdal1960}%
  \BibitemOpen
  \bibfield  {author} {\bibinfo {author} {\bibfnamefont {A.~B.}\ \bibnamefont
  {Migdal}},\ }\href@noop {} {\bibfield  {journal} {\bibinfo  {journal} {JETP}\
  }\textbf {\bibinfo {volume} {10}},\ \bibinfo {pages} {176} (\bibinfo {year}
  {1960})}\BibitemShut {NoStop}%
\bibitem [{\citenamefont {Lyubashevsky}\ \emph {et~al.}(2025)\citenamefont
  {Lyubashevsky} \emph {et~al.}}]{lyubashevsky2025}%
  \BibitemOpen
  \bibfield  {author} {\bibinfo {author} {\bibfnamefont {D.~E.}\ \bibnamefont
  {Lyubashevsky}} \emph {et~al.},\ }\href
  {https://doi.org/10.1088/1674-1137/ad8d4d} {\bibfield  {journal} {\bibinfo
  {journal} {Chin. Phys. C}\ }\textbf {\bibinfo {volume} {48}} (\bibinfo {year}
  {2025})}\BibitemShut {NoStop}%
\bibitem [{\citenamefont {Randrup}\ and\ \citenamefont
  {Vogt}(2009)}]{randrup2009}%
  \BibitemOpen
  \bibfield  {author} {\bibinfo {author} {\bibfnamefont {J.}~\bibnamefont
  {Randrup}}\ and\ \bibinfo {author} {\bibfnamefont {R.}~\bibnamefont {Vogt}},\
  }\href {https://doi.org/10.1103/PhysRevC.80.024601} {\bibfield  {journal}
  {\bibinfo  {journal} {Phys. Rev. C}\ }\textbf {\bibinfo {volume} {80}},\
  \bibinfo {pages} {024601} (\bibinfo {year} {2009})}\BibitemShut {NoStop}%
\bibitem [{\citenamefont {Verbeke}\ \emph {et~al.}(2018)\citenamefont {Verbeke}
  \emph {et~al.}}]{verbeke2018}%
  \BibitemOpen
  \bibfield  {author} {\bibinfo {author} {\bibfnamefont {J.~M.}\ \bibnamefont
  {Verbeke}} \emph {et~al.},\ }\href
  {https://doi.org/10.1016/j.cpc.2017.09.006} {\bibfield  {journal} {\bibinfo
  {journal} {Comput. Phys. Commun.}\ }\textbf {\bibinfo {volume} {222}},\
  \bibinfo {pages} {263} (\bibinfo {year} {2018})}\BibitemShut {NoStop}%
\bibitem [{\citenamefont {Dössing}\ and\ \citenamefont
  {Randrup}(1985{\natexlab{a}})}]{dossing_I_1985}%
  \BibitemOpen
  \bibfield  {author} {\bibinfo {author} {\bibfnamefont {T.}~\bibnamefont
  {Dössing}}\ and\ \bibinfo {author} {\bibfnamefont {J.}~\bibnamefont
  {Randrup}},\ }\href {https://doi.org/10.1016/0375-9474(85)90178-2} {\bibfield
   {journal} {\bibinfo  {journal} {Nucl. Phys. A}\ }\textbf {\bibinfo {volume}
  {433}},\ \bibinfo {pages} {215} (\bibinfo {year}
  {1985}{\natexlab{a}})}\BibitemShut {NoStop}%
\bibitem [{\citenamefont {Dössing}\ and\ \citenamefont
  {Randrup}(1985{\natexlab{b}})}]{dossing_II_1985}%
  \BibitemOpen
  \bibfield  {author} {\bibinfo {author} {\bibfnamefont {T.}~\bibnamefont
  {Dössing}}\ and\ \bibinfo {author} {\bibfnamefont {J.}~\bibnamefont
  {Randrup}},\ }\href {https://doi.org/10.1016/0375-9474(85)90179-4} {\bibfield
   {journal} {\bibinfo  {journal} {Nucl. Phys. A}\ }\textbf {\bibinfo {volume}
  {433}},\ \bibinfo {pages} {280} (\bibinfo {year}
  {1985}{\natexlab{b}})}\BibitemShut {NoStop}%
\bibitem [{\citenamefont {Vogt}\ and\ \citenamefont
  {Randrup}(2013)}]{vogt2013}%
  \BibitemOpen
  \bibfield  {author} {\bibinfo {author} {\bibfnamefont {R.}~\bibnamefont
  {Vogt}}\ and\ \bibinfo {author} {\bibfnamefont {J.}~\bibnamefont {Randrup}},\
  }\href {https://doi.org/10.1103/PhysRevC.87.044602} {\bibfield  {journal}
  {\bibinfo  {journal} {Phys. Rev. C}\ }\textbf {\bibinfo {volume} {87}},\
  \bibinfo {pages} {044602} (\bibinfo {year} {2013})}\BibitemShut {NoStop}%
\bibitem [{\citenamefont {Randrup}\ and\ \citenamefont
  {Vogt}(2014)}]{randrup2014}%
  \BibitemOpen
  \bibfield  {author} {\bibinfo {author} {\bibfnamefont {J.}~\bibnamefont
  {Randrup}}\ and\ \bibinfo {author} {\bibfnamefont {R.}~\bibnamefont {Vogt}},\
  }\href {https://doi.org/10.1103/PhysRevC.89.044601} {\bibfield  {journal}
  {\bibinfo  {journal} {Phys. Rev. C}\ }\textbf {\bibinfo {volume} {89}},\
  \bibinfo {pages} {044601} (\bibinfo {year} {2014})}\BibitemShut {NoStop}%
\bibitem [{\citenamefont {Randrup}(2022)}]{randrup2022}%
  \BibitemOpen
  \bibfield  {author} {\bibinfo {author} {\bibfnamefont {J.}~\bibnamefont
  {Randrup}},\ }\href {https://doi.org/10.1103/PhysRevC.106.L051601} {\bibfield
   {journal} {\bibinfo  {journal} {Phys. Rev. C}\ }\textbf {\bibinfo {volume}
  {106}},\ \bibinfo {pages} {L051601} (\bibinfo {year} {2022})}\BibitemShut
  {NoStop}%
\bibitem [{\citenamefont {Scamps}\ \emph {et~al.}(2023)\citenamefont {Scamps}
  \emph {et~al.}}]{scamps2023}%
  \BibitemOpen
  \bibfield  {author} {\bibinfo {author} {\bibfnamefont {G.}~\bibnamefont
  {Scamps}} \emph {et~al.},\ }\href
  {https://doi.org/10.1103/PhysRevC.108.L061602} {\bibfield  {journal}
  {\bibinfo  {journal} {Phys. Rev. C}\ }\textbf {\bibinfo {volume} {108}},\
  \bibinfo {pages} {L061602} (\bibinfo {year} {2023})}\BibitemShut {NoStop}%
\bibitem [{\citenamefont {Chen}\ \emph {et~al.}(2023)\citenamefont {Chen} \emph
  {et~al.}}]{chen2023}%
  \BibitemOpen
  \bibfield  {author} {\bibinfo {author} {\bibfnamefont {J.}~\bibnamefont
  {Chen}} \emph {et~al.},\ }\href
  {https://doi.org/10.1051/epjconf/202328404006} {\bibfield  {journal}
  {\bibinfo  {journal} {EPJ Web Conf.}\ }\textbf {\bibinfo {volume} {284}},\
  \bibinfo {pages} {04006} (\bibinfo {year} {2023})}\BibitemShut {NoStop}%
\bibitem [{\citenamefont {Kadmensky}\ \emph
  {et~al.}(2024{\natexlab{b}})\citenamefont {Kadmensky} \emph
  {et~al.}}]{kadmensky_ran2024}%
  \BibitemOpen
  \bibfield  {author} {\bibinfo {author} {\bibfnamefont {S.~G.}\ \bibnamefont
  {Kadmensky}} \emph {et~al.},\ }\href
  {https://doi.org/10.1134/S1062873824707384} {\bibfield  {journal} {\bibinfo
  {journal} {Bull. Russ. Acad. Sci. Phys.}\ }\textbf {\bibinfo {volume} {88}},\
  \bibinfo {pages} {1249} (\bibinfo {year} {2024}{\natexlab{b}})}\BibitemShut
  {NoStop}%
\bibitem [{\citenamefont {Scamps}(2024)}]{scamps2024}%
  \BibitemOpen
  \bibfield  {author} {\bibinfo {author} {\bibfnamefont {G.}~\bibnamefont
  {Scamps}},\ }\href {https://doi.org/10.1103/PhysRevC.109.L011602} {\bibfield
  {journal} {\bibinfo  {journal} {Phys. Rev. C}\ }\textbf {\bibinfo {volume}
  {109}},\ \bibinfo {pages} {L011602} (\bibinfo {year} {2024})}\BibitemShut
  {NoStop}%
\end{thebibliography}%

\end{document}